\documentclass[12pt]{article}
\usepackage{amsmath,amsfonts,amssymb}
\usepackage{graphicx}
\usepackage[a4paper]{geometry}
\usepackage{subfigure}
\usepackage{url}

\newtheorem{thm}{Theorem}
\newtheorem{defi}{Definition}

\newtheorem{prop}{Proposition}

\newtheorem{assumption}{Assumption}
\newtheorem{remark}{Remark}
\newtheorem{example}{Example}

\newcommand{\bmath}[1]{\mbox{\boldmath{$#1$}}}

\begin{document}

\title{Incorporating fat tails in financial models using entropic divergence measures}

\author{
        Santanu Dey \  \  \  \  \     Sandeep Juneja                    \\
        \\
       Tata Institute of Fundamental Research\\
        Mumbai, India       \\
        \\
       dsantanu@tcs.tifr.res.in \  \  \  \  juneja@tifr.res.in                       \\
            }

\maketitle

\setcounter{page}{1}
\sloppy
\vspace{40mm}

\begin{abstract}
In the existing financial literature, entropy based ideas have been proposed in portfolio optimization,
in model calibration for options pricing as well as in ascertaining a pricing measure in incomplete markets.
The abstracted problem corresponds to finding a probability measure that minimizes the relative entropy (also called $I$-divergence)
with respect to a known measure  while it satisfies certain moment  constraints on functions of underlying assets.
In this paper, we show that under  $I$-divergence,
the optimal solution  may not exist when the underlying assets have fat tailed distributions, ubiquitous in financial practice.
We note that this drawback may be corrected if `polynomial-divergence' is used.
This divergence can be seen to be equivalent to the well known (relative) Tsallis or (relative) Renyi entropy.
We discuss existence and uniqueness issues related to this new optimization problem as well as  the nature of the optimal solution under different objectives.
We also identify the optimal solution structure under $I$-divergence as well as polynomial-divergence when the associated constraints include those on marginal
distribution of  functions of underlying assets.
These results are applied to a simple problem of model calibration to options prices  as well as  to portfolio modeling in Markowitz framework,
where we note that a reasonable view that a  particular portfolio of assets has heavy tailed losses  may lead to fatter and more reasonable tail distributions of all assets.
\end{abstract}

\newpage
\section{Introduction}

Entropy based ideas have found a number of  popular applications in finance over the last two decades.
A key application involves  portfolio optimization where  these are used (see, e.g.,  Meucci \cite{fully:flexible})
to arrive at a `posterior' probability measure that is closest to the specified `prior' probability measure
and satisfies expert views modeled as constraints on certain moments associated with the posterior probability measure.
Another important application  involves calibrating the risk neutral probability measure used for pricing options
(see, e.g., Buchen and Kelly \cite{buch:kelly}, Stutzer \cite{stutzer}, Avellaneda et al. \cite{avel3}).
Here entropy based ideas are used to arrive at a probability measure that correctly prices given liquid options
while again being closest to a specified  `prior' probability measure.

In these works, relative entropy or $I$-divergence is used as a measure of distance between two probability measures.
One advantage of $I$-divergence is that under mild conditions, the posterior probability measure exists and has an
elegant representation when the underlying model corresponds to a distribution of light tailed random variables (that is, when the moment generating function exists in a neighborhood around zero).
However, we note that this is no longer true when the underlying random variables may be fat-tailed,
as is often the case in finance and insurance settings.
One of our key contributions is to observe that when probability distance measures corresponding to `polynomial-divergence'
(defined later)  are used in place of $I$-divergence, under technical conditions, the posterior probability measure exists
and has an elegant representation even when the underlying random variables may be fat-tailed.
Thus, this provides a reasonable way to incorporate restrictions in the presence of fat tails.
Our another main contribution is that we devise a methodology  to arrive at a posterior probability measure when the constraints on
this measure are of a general nature that include  specification of marginal distributions of functions of
underlying random variables. For instance, in portfolio optimization settings, an expert may have a view that
certain index of stocks has a fat-tailed t-distribution and is looking for a posterior distribution that
satisfies this requirement while being closest to a prior model that may, for instance,  be based on historical data.

\subsection{Literature Overview}

The  evolving literature on  updating models for portfolio optimization  builds upon  the pioneering work of Black and Litterman \cite{black:litterman} (BL).
BL  consider variants of Markowitz's model  where the subjective views of portfolio managers are used as constraints to update models of the market using ideas
from Bayesian analysis. Their work focuses on Gaussian framework with views restricted to  linear combinations of expectations of returns from different securities.
Since then a number of variations and improvements have been suggested
(see, e.g., \cite{mina:xiao}, \cite{least:disc} and \cite{qian:gorman}).
Recently, Meucci \cite{fully:flexible} proposed `entropy pooling' where the original model can involve general distributions and views can be on
any characteristic of the probability model.
Specifically, he focuses on approximating the original distribution of asset returns by a discrete one generated via Monte-Carlo sampling (or from data).
Then a convex programming problem is solved that adjusts  weights to these samples so that they minimize the $I$-divergence from the original sampled distribution
while satisfying the view constraints.
These samples with updated weights are then used for portfolio optimization.
Earlier,  Avellaneda et al. \cite{avel3} used similar weighted Monte Carlo methodology  to calibrate asset pricing models to market data
(see also Glasserman and Yu \cite{glasserman:yu}).
Buchen and Kelly in \cite{buch:kelly} and Stutzer in \cite{stutzer} use the entropy approach to calibrate one-period
asset pricing models by selecting a pricing measure that correctly prices  a set of benchmark instruments while minimizing
$I$-divergence from  a prior specified model, that may, for instance be estimated from historical data (see also the recent survey article \cite{kitamura:stutzer}).

Zhou et.al in \cite{cf3} consider a  statistical learning framework for  estimating default probabilities of private firms over a fixed horizon of time,
given market data on various `explanatory variables' like stock prices, financial ratios and other economic indicators.
They  use the entropy approach to calibrate the default probabilities (see also Friedman and Sandow \cite{cf4}).


Another popular application of entropy based ideas to finance  is in valuation of non-hedgeble payoffs in incomplete markets.
In an incomplete market there may exist many probability measures equivalent to the physical probability  measure under which
discounted price processes of the underlying assets are martingales.
Fritelli (\cite{fritelli}) proposes using a probability measure that minimizes the $I$-divergence from the physical probability measure
for pricing purposes (he calls this the minimal entropy martingale measure or MEMM). Here, the underlying financial model corresponds to a continuous time stochastic process.  Cont and Tankov (\cite{cont:tankov}) consider, in addition, the problem of incorporating calibration constraints into MEMM for exponential Levy processes
(see also Kallsen \cite{kallsen}). Jeanblanc et al. (\cite{fq}) propose using polynomial-divergence
(they call it $f^q$-divergence distance) instead of $I$-divergence and obtain necessary and sufficient conditions for existence of an EMM with minimal
polynomial-divergence from the physical measure for exponential Levy processes.
They also characterize the form of density of the optimal measure.
Goll and Ruschendorf (\cite{goll}) consider general $f$-divergence in a general semi-martingale setting to single out an EMM which also satisfies calibration constraints.

A brief historical perspective on related entropy based literature may be in order:
This concept  was first introduced by Gibbs in the framework of
classical theory of thermodynamics (see \cite{gibbs}) where entropy was defined as a  measure of \textit{disorder} in thermodynamical systems.
Later, Shannon  \cite{Shannon:one} (see also \cite{khin:chin})  proposed that entropy could be interpreted as a measure of
\textit{missing information} of a random system.
Jaynes \cite{jns:one}, \cite{jns:two} further developed this idea in the framework of statistical inferences and gave the mathematical formulation of
\textit{principle of maximum entropy} (PME),
which states that given partial information/views or constraints about an unknown probability distribution, among all the
distributions that satisfy these restrictions, the distribution that maximizes the entropy is the one that is least prejudiced,
in the sense of being minimally committal to missing information.
When a prior probability distribution, say,  $\mu$ is given, one can extend above principle to \textit{principle of minimum cross entropy} (PMXE),
which states that among all probability measures which satisfy a given set of constraints, the one with minimum relative entropy (or the $I$-divergence) with
respect to $\mu$, is the one that is maximally committal to the prior $\mu$.
 See \cite{jn:kapur} for numerous applications of PME and PMXE in
diverse fields of science and engineering. See \cite{csiszar:2008}, \cite{s:abe}, \cite{dos:santos}, \cite{golshani:pasha} and \cite{jizba:arimitsu}
for axiomatic justifications for Renyi and Tsallis entropy.
For information theoretic import of $f$-divergence and its relation to utility maximization see Friedman et.al. \cite{cf1}, Slomczynski and Zastawniak \cite{zasta}.

\subsection{Our Contributions}

In this article,  we restrict attention to examples related to portfolio optimization and model calibration
and  build upon ideas proposed by Avellaneda et al. \cite{avel3},  Buchen and Kelly \cite{buch:kelly}, Meucci \cite{fully:flexible} and others.
We first note  the well known result  that for views expressed as finite number of moment constraints,
the optimal solution to the $I$-divergence minimization can be  characterized as a  probability measure
obtained by suitably exponentially twisting the original measure.
This measure is known in literature as the Gibbs measure and our analysis is based on the well known
ideas involved in Gibbs conditioning principle  (see, for instance, \cite{dembo:zeit}).
As mentioned earlier,  such a characterization may fail when the underlying distributions are fat-tailed
in the sense that their moment generating function does not exist in
some neighborhood of origin.
We show that one reasonable way to get a good change of measure that incorporates views\footnote{In this article, we often use `views' or `constraints' interchangeably}
in this setting is by replacing $I$-divergence by a suitable `polynomial-divergence' as an objective in our optimization problem.
We characterize the optimal solution  in this setting,
and prove its uniqueness under  technical conditions.
Our definition of polynomial-divergence is a special case of a more general concept of
\textit{f-divergence} introduced by Csiszar in \cite{csiszar1:1967}, \cite{csiszar2:1967} and \cite{csiszar:1972}.
Importantly,  polynomial-divergence is  monotonically increasing function of the well known relative
\textit{Tsallis Entropy} \cite{tsallis:one} and relative \textit{Renyi Entropy} \cite{alfred:renyi},
and moreover, under appropriate limit converges to $I$-divergence.

As indicated earlier, we also consider the case where the expert views may specify marginal probability distribution of functions of random variables involved.
We show that such views, in addition to views on moments of functions of underlying random variables are easily incorporated. In particular,
under technical conditions, we characterize the optimal solution with these general constraints, when the objective may be
$I$-divergence or polynomial-divergence and show the uniqueness of the resulting optimal probability measure in each case.

As an illustration,  we apply these results to portfolio modeling in Markowitz framework
where the returns from a finite number of assets have a multivariate Gaussian distribution and expert view is that
a certain portfolio of returns  is fat-tailed. We show that in the resulting probability measure, under mild conditions, all assets are similarly fat-tailed.
Thus, this becomes a reasonable way to incorporate realistic tail behavior in a portfolio of assets.
Generally speaking,  the proposed approach may be useful in better risk management by building conservative tail views in mathematical models.

Note that a key reason to propose polynomial-divergence is that it provides a tractable and elegant way to arrive at a reasonable updated distribution
close to the given prior distribution while  incorporating constraints and views even when fat-tailed distributions are involved.
It is natural to try to understand the influence of  the choice of objective function on the resultant optimal probability measure.
We address this issue for a simple example where we compare the three reasonable objectives:
the total variational distance, $I$-divergence and polynomial-divergence.
We discuss the differences in the resulting solutions.
To shed further light on this, we also observe that when views are expressed as constraints on probability values of disjoint sets,
the optimal solution is the same in all three cases. Furthermore, it has a simple representation.
We also conduct numerical experiments on practical examples to validate the proposed methodology.

\subsection{Organization}

In Section 2, we outline the mathematical framework and characterize the optimal probability measure
that minimizes the $I$-divergence with respect to the original probability measure subject to views expressed as  moment constraints of specified functions.
In Section 3, we show through an example that $I$-divergence may be inappropriate objective function in the presence of fat-tailed distributions.
We then define polynomial-divergence and characterize the optimal probability measure that minimizes this divergence subject to constraints on moments.
The uniqueness of the optimal measure, when it exists, is proved under technical assumptions. We also discuss existence of the solution in a simple setting. When under the proposed methodology a solution does not exist, we also propose a weighted least squares based modification that finds a reasonable perturbed solution to the problem.
In Section 4, we extend the methodology to incorporate views on marginal distributions of some random variables,
along with views on moments of functions of underlying random variables.
We characterize the optimal probability measures that minimize $I$-divergence and polynomial-divergence in this setting
and prove the uniqueness of the optimal measure when it exists.
In Section 5, we apply our results to  the portfolio problem in the Markowitz framework and develop explicit expressions for the posterior probability measure.
We also show how a view that a portfolio of assets has a `regularly varying' fat-tailed distribution renders a similar fat-tailed marginal distribution to all assets correlated to this portfolio.
Section 6 is devoted to comparing qualitative differences on a simple example in the resulting optimal probability measures when the objective function is
$I$-divergence, polynomial-divergence and total variational distance.
In this section, we also note that when views are on probabilities of disjoint sets, all three objectives give identical results.
We numerically test our proposed algorithms on practical examples in Section 7.
Finally, we end in  Section 8 with a brief conclusion.
All but the simplest proofs are relegated to the Appendix.

We thank the anonymous referee for bringing Friedman et al \cite{cf4} to our notice.
This article also considers $f$-divergence and in particular, polynomial-divergence (they refer to
an equivalent quantity as $u$-entropy)
in the setting of
univariate power-law distribution which includes Pareto and Skewed generalized $t$-distribution.
It motivates the use of polynomial-divergence through utility maximization considerations
and  develops practical and robust calibration technique for univariate asset return densities.

\section{Incorporating Views using $I$-Divergence}

Some notation and basic concepts are needed to support our analysis.
Let $(\Omega,\mathcal{F},\mu)$ denote the underlying probability space.
Let $\mathcal{P}$ be the set of all probability measures on $(\Omega,\mathcal{F})$.
For any $\nu\in\mathcal{P}$ the \textit{relative entropy} of $\nu$ w.r.t $\mu$
 or \textit{$I$-divergence} of $\nu$ w.r.t $\mu$ (equivalently, the \textit{Kullback-Leibler distance}\footnote{though this is not a distance in the strict sense of a metric})
is defined as
$$\displaystyle  D(\nu\mid\mid\mu):= \int \log\left(\frac{d\nu}{d\mu}\right)\,d\nu  $$
 if $\nu$ is absolutely continuous with respect to $\mu$ and
 $\ log(\frac{d\nu}{d\mu})$ is integrable.
 $D(\nu\mid\mid\mu)=+\infty$ otherwise. See, for instance \cite{thomas:cover},
 for concepts related to relative entropy.

Let $\mathcal{P}(\mu)$ be the set of all probability measures which are absolutely continuous w.r.t. $\mu$, $\psi:\Omega\rightarrow\mathbb{R}$ be a  measurable function
such that $\int |\psi |e^\psi d \mu < \infty$, and let
$$\Lambda(\psi):=\log\int e^\psi\,d\mu \ \in(-\infty,+\infty]$$
denote the logarithmic moment generating function of $\psi$ w.r.t \ $\mu$.
Then it is well known  that
$$\Lambda(\psi)
=\sup_{\nu \in \mathcal{P}(\mu)}
\{\int\psi\,d\nu-D(\nu\mid\mid\mu)\}.$$
Furthermore, this supremum  is attained at $\nu^*$ given by:
\begin{equation} \label{eqn:opt_measure}
 \frac{d\nu^*}{d\mu}=\frac{e^\psi}{\int e^\psi\,d\mu}.
\end{equation}
(see, for instance, \cite{thomas:cover}, \cite{jn:kapur}, \cite{dup:eli}).

In our optimization problem
we look for a probability measure $\nu \in \mathcal{P}(\mu)$ that minimizes the $I$-divergence w.r.t. $\mu$.
We restrict our search to probability measures that satisfy moment constraints
 $\int g_i\,d\nu \geq c_i,$
 and/or
$\int g_i\,d\nu = c_i,$
where each $g_i$ is a measurable function.
For instance, views on probability of certain sets can be modeled by setting\ $g_i$'s as indicator functions of those sets. If our underlying space supports
random variables $(X_1, \ldots, X_n)$ under the probability measure $\mu$,
one may set $g_i=f_i(X_1, \ldots, X_n)$ so that the associated constraint is on the expectation of these functions.

Formally, our optimization problem ${\bf O_1}$ is:

\begin{equation} \label{prob1}
\min_{\nu \in \mathcal{P}(\mu)}
\int \log\left(\frac{d\nu}{d\mu}\right)\,d\nu
\end{equation}
subject to the constraints:
\begin{equation} \label{eqn:constr_1}
\int g_i\,d\nu \geq c_i,
\end{equation}
for $i=1,\ldots,k_1$ and
\begin{equation} \label{eqn:constr_2}
\int g_i\,d\nu = c_i,
\end{equation}
for $i=k_1+1, \ldots, k$. Here $k_1$ can take any value between 0 and $k$.

The solution to this is characterized by the following assumption:


\begin{assumption}
\label{assm1}
There exist $\lambda_i \geq 0$ for $i=1, \ldots, k_1$,  and
$\lambda_{k_1+1},...,\lambda_k\in\mathbb{R}$
such that
$$\int e^{\sum_i\lambda_ig_i}\,d\mu < \infty$$
and the probability measure\ $\nu^0$ given by
\begin{equation}
\label{form}
\nu^0(A)= \int_A \frac{e^{\sum_i\lambda_ig_i}\,d\mu}{\int e^{\sum_i\lambda_ig_i}\,d\mu}
\end{equation}
for all $A \in \mathcal{F}$ satisfies the constraints (\ref{eqn:constr_1}) and (\ref{eqn:constr_2}).
Furthermore, the complementary slackness conditions
$$\lambda_i (c_i -\int g_i\,d\nu)=0,$$
hold for $i=1, \ldots, k_1$.
\end{assumption}
The following theorem follows:
\begin{thm}\label{klmom}
Under Assumption \ref{assm1}, $\nu^0$ is an optimal solution to ${\bf O_1}$.
\end{thm}

This theorem is well known and a proof using Lagrange multiplier method can be found in \cite{thomas:cover}, \cite{dup:eli}, \cite{buch:kelly} and \cite{avel1}.
For completeness we sketch the proof below.

\vspace{0.1in}

\noindent \textbf{Proof of Theorem \ref{klmom}:}
${\bf O_1}$ is equivalent to maximizing $-D(\nu\mid\mid\mu)=-\int log(\frac{d\nu}{d\mu})\,d\nu$ subject to the constraints (\ref{eqn:constr_1})
and (\ref{eqn:constr_2}).
The Lagrangian for the above maximization problem is:\\
\begin{eqnarray*}
\mathcal{L}&=&\sum_i\lambda_i \left(\int g_i\,d\nu - c_i \right) + (-D(\nu\mid\mid\mu))\\
           &=&\int\psi\,d\nu -D(\nu\mid\mid\mu) - \sum_i\lambda_i c_i,
\end{eqnarray*}
where  $\psi=\sum_i\lambda_ig_i$.
Then by (\ref{eqn:opt_measure}) and the preceding discussion, it follows
that  $\nu^0$  maximizes \ $\mathcal{L}$.
By Lagrangian duality, due to Assumption~\ref{assm1}, $\nu^0$ also solves ${\bf O_1}$.
$\Box$

Note that to obtain the optimal distribution by formula (\ref{form}), we must solve the constraint equations for the Lagrange multipliers
$\lambda_1, \lambda_2,\ldots, \lambda_k$. The constraint equations with its explicit dependence on $\lambda_i$'s can be written as:
\begin{equation}
\label{eqn:221}
\frac{\int g_j e^{\sum_i\lambda_ig_i}\,d\mu}{\int e^{\sum_i\lambda_ig_i}\,d\mu}=c_j\,\,\,\text{for}\,\,j=1,2,\ldots,k.
\end{equation}
This is a set of $k$ nonlinear equations in $k$ unknowns $\lambda_1,\lambda_2,\ldots,\lambda_k$,
and typically would require numerical procedures for solution.
It is easy to see that if the constraint equations are not consistent then no solution exists.
For a sufficient condition for existence of a solution,
see Theorem~3.3 of \cite{csiszar:1975}.
On the other hand, when a solution does exist, it is helpful for applying numerical procedures, to know if it is unique.
It can be shown that the Jacobian matrix of the set of equation (\ref{eqn:221}) is given by the variance-covariance matrix of
$g_1,g_2,\ldots,g_k$ under the measure given by (\ref{form}).
The last mentioned variance-covariance matrix is also equal to the Hessian of the following function
$$G(\lambda_1,\lambda_2,\ldots,\lambda_k):=\int e^{\sum_i\lambda_ig_i}\,d\mu -\sum_i\lambda_i c_i.$$
For details,
see \cite{buch:kelly} or \cite{avel1}.
It is easily  checked that (\ref{eqn:221}) is same as
\begin{equation}
\left(\frac{\partial G}{\partial\lambda_1},\frac{\partial G}{\partial\lambda_2},\ldots,\frac{\partial G}{\partial\lambda_k} \right)=0\,.
\end{equation}
It follows that if no non-zero linear combination of $g_1,g_2,\ldots,g_k$ has zero variance under the measure given by (\ref{form}),
then G is strictly convex and if a solution to (\ref{eqn:221}) exists, it is unique.
It also follows that instead of employing a root-search procedure to solve (\ref{eqn:221}) for $\lambda_i$'s,
one may as well find a local minima (which is also global) of the function $G$ numerically.
We end this section with a simple example.

\begin{example}  {\em
Suppose that under $\mu$,
random variables $X=(X_1, \ldots, X_n)$ have a multivariate normal distribution
$N(a, \Sigma)$, that is,  with mean $a\in\mathbb{R}^n$ and variance covariance matrix
$\Sigma\in\mathbb{R}^{n\times n}$. If constraints  correspond to
their mean vector being equal to $\hat{a}$,
then this can be achieved by a new probability measure $\nu^0$ obtained
by exponentially twisting $\mu$ by a vector $\lambda \in\mathbb{R}^n$ such that
$$\lambda =  (\Sigma^{-1})^T(\hat{a}-a).$$
Then, under $\nu^0$, $X$ is  $N(\tilde{a}, \Sigma)$ distributed.}
\end{example}


\section{Incorporating Views using Polynomial-Divergence}
\label{polynomial:div}
In this section, we first note through a simple example involving a fat-tailed distribution that optimal solution under $I$-divergence
may not exist in certain settings. In fact, in this simple setting,
one can obtain a solution that is arbitrarily close to the original distribution in the sense of $I$-divergence.
However, the form of such solutions may be inelegant and not  reasonable as a model in financial settings.
This motivates the use of other notions of distance between probability measures  as  objectives such as $f$-divergence (introduced by Csiszar~\cite{csiszar:1972}).
We first define general $f$-divergence and later concentrate on the case where $f$ has the form $f(u)=u^{\beta+1}$, $\beta>0$.
We refer to this as polynomial-divergence and note its relation with relative Tsallis entropy, relative Renyi entropy and $I$-divergence.
We then characterize the optimal solution under polynomial-divergence.
To provide greater insight into nature of this solution, we explicitly solve a few examples in a
simple setting  involving a single random variable and a single moment constraint.
We also note that in some settings under polynomial-divergence as well an optimal solution may not exist.  With a view to arriving at a pragmatic remedy,  we  then describe a weighted  least squares based methodology to  arrive at a solution by perturbing certain parameters by a minimum amount.

\subsection{Polynomial Divergence}
\begin{example} \label{exam:1}  {\em
Suppose that under $\mu$, non-negative
random variable $X$ has a Pareto distribution with
probability density function
$$f(x) = \frac{\alpha-1}{(1+x)^{\alpha}}\,\,,\,x\geq0,\,\,\alpha>2.$$
The mean under this pdf equals $1/(\alpha-2)$. Suppose the view is that
the mean should  equal  $c > 1/(\alpha-2)$.
It is well known and easily checked that
$$\frac{\int x e^{\lambda x} f(x) dx}{\int e^{\lambda x} f(x) dx}$$
is an increasing function of $\lambda$ that equals $\infty$ for $\lambda >0$.
Hence, Assumption~1, does not hold for this example.
Similar difficulty arises with other fat-tailed distributions such as  log-normal and $t$-distribution.}
\end{example}

To shed further light on Example~\ref{exam:1}, for $M>0$,  consider a probability distribution
$${f}_{\lambda}(x) = \frac{\exp( \lambda x) f(x) I_{[0,M]}(x)}{\int_0^M \exp( \lambda x) f(x) dx}.$$
Where $I_A(\cdot)$ denote the indicator function of the set $A$, that is, $I_A(x)=1$ if $x\in A$ and $0$ otherwise.
Let  $\lambda_M$ denote the solution to
$$\int_0^{\infty}x {f}_{\lambda}(x) dx = c \,\,(> 1/(\alpha-2)).$$
\begin{prop} \label{prop:001}
The sequence $\{\lambda_M \}$ and the the $I$-divergence  $\int \log \left (\frac{{f}_{\lambda_M}(x)}{f(x)} \right ) f_{\lambda_M}(x)dx$
both converge to zero as  $M \rightarrow \infty$.
\end{prop}

Solutions such as ${f}_{\lambda_M}$  above are typically not representative of many applications, motivating the need to
have alternate methods to arrive at reasonable posterior measures that are close to $\mu$
and satisfy constraints such as
(\ref{eqn:constr_1}) and (\ref{eqn:constr_2}) while not requiring  that the optimal solution be obtained using exponential twisting.

We now address this issue using polynomial-divergence. We first define $f$-divergence introduced by Csiszar (see \cite{csiszar1:1967}, \cite{csiszar2:1967}
and \cite{csiszar:1972}). 

\begin{defi}
Let $f:(0,\infty)\rightarrow\mathbb{R}$ be a strictly convex function. The $f$-divergence of a probability measure $\nu$  w.r.t. another probability measure $\mu$
equals
$$I_f(\nu\mid\mid\mu):=\int f\left(\frac{d\nu}{d\mu}\right)\,d\mu $$
if $\nu$  is absolutely continuous  and
$f(\frac{d\nu}{d\mu})$ is integrable w.r.t. $\mu$. Otherwise we set
$I_f(\nu\mid\mid\mu)=\infty$.
\end{defi}

Note that $I$-divergence corresponds to the case $f(u)=u \log u$. Other popular examples of  $f$ include
$$f(u)=-\log u,\ f(u)=u^{\beta+1},\ \beta>0,\ f(u)=e^u.$$
In this section we consider $f(u)=u^{\beta+1},\ \beta>0$ and refer to the resulting $f$-divergence as {\em polynomial-divergence}. That is, we focus on
$$I_{\beta}(\nu\mid\mid\mu):=\int\left(\frac{d\nu}{d\mu}\right)^\beta\,d\nu=\int\left(\frac{d\nu}{d\mu}\right)^{\beta+1}\,d\mu.$$
It is easy to see using Jensen's inequality that
$$\min_{\nu \in \mathcal{P}(\mu)}\int\left(\frac{d\nu}{d\mu}\right)^{\beta+1}\,d\mu$$
is achieved by $\nu=\mu$.


Our optimization problem ${\bf O_2}(\beta)$ may be stated as:
\begin{equation}
\min_{\nu \in \mathcal{P}(\mu)}
\int\left(\frac{d\nu}{d\mu}\right)^{\beta+1}\,d\mu
\end{equation}
subject to (\ref{eqn:constr_1}) and (\ref{eqn:constr_2}).
Minimizing polynomial-divergence
can also be motivated through utility maximization considerations.
See \cite{cf1} for further details.

\begin{remark}{Relation with Relative Tsallis Entropy and Relative Renyi Entropy:}\em{
Let $\alpha$ and $\gamma$ be a positive real numbers.
The relative Tsallis entropy with index $\alpha$ of $\nu$  w.r.t. $\mu$
equals
$$S_{\alpha}(\nu\mid\mid\mu):=\int\frac{\left(\frac{d\nu}{d\mu}\right)^{\alpha}-1}{\alpha}d\nu,$$
if $\nu$  is absolutely continuous w.r.t.  $\mu$ and the integral is finite.
Otherwise,
$S_{\alpha}(\nu\mid\mid\mu)=\infty$. See, e.g., \cite{tsallis:one}.
The relative Renyi entropy of order $\gamma$ of $\nu$  w.r.t. $\mu$
equals
$$H_{\gamma}(\nu\mid\mid\mu):=\frac{1}{\gamma-1}\log\left(\int \left(\frac{d\nu}{d\mu}\right)^{\gamma-1}\,d\nu\right)\,\,\,\text{when}\,\, \gamma\neq1$$
and
$$H_{1}(\nu\mid\mid\mu):=\int \log\left(\frac{d\nu}{d\mu}\right)\,d\nu,$$
if $\nu$  is absolutely continuous w.r.t $\mu$ and the respective integrals are finite.
Otherwise, $H_{\gamma}(\nu\mid\mid\mu)=\infty$ (see, e.g, \cite{alfred:renyi}).

It can be shown that as $\gamma\rightarrow 1\,\,,\,H_{\gamma}(\nu\mid\mid\mu)\rightarrow H_{1}(\nu\mid\mid\mu)=D(\nu\mid\mid\mu)$\,\, and as
$\alpha\rightarrow 0\,\,,\,S_{\alpha}(\nu\mid\mid\mu)\rightarrow D(\nu\mid\mid\mu).$
Also, following relations are immediate consequences of the above definitions:
$$I_{\beta}(\nu\mid\mid\mu)=1+\beta S_{\beta}(\nu\mid\mid\mu),$$
$$I_{\beta}(\nu\mid\mid\mu)=e^{\beta H_{\beta+1}((\nu\mid\mid\mu)}$$
and
$$\lim_{\beta\to 0}\frac{I_{\beta}(\nu\mid\mid\mu)-1}{\beta}=D(\nu\mid\mid\mu).$$

Thus, polynomial-divergence is a strictly increasing function of both relative Tsallis entropy and relative Renyi entropy.
Therefore minimizing polynomial-divergence is equivalent to minimizing relative Tsallis entropy or relative Renyi entropy.
}
\end{remark}

In the following assumption we specify the form of solution to ${\bf O_2}(\beta)$:


\begin{assumption}
\label{assm2}
There exist $\lambda_i \geq 0$ for $i=1, \ldots, k_1$,  and
$\lambda_{k_1+1},...,\lambda_k\in\mathbb{R}$
such that
\begin{equation}\label{ineq:200}
1+\beta \sum_{l=1}^k\lambda_lg_l \geq 0\,\,\,a.e. (\mu).
\end{equation}
\begin{equation}\label{ineq:201}
\int\left(1+ \beta \sum_{l=1}^k\lambda_lg_l\right)^{1+ \frac{1}{\beta}}\,d\mu < \infty,
\end{equation}
and the probability measure
$\nu^1$ given by
\begin{equation}
\label{form2}
\nu^1(A)= \int_A \frac{(1+ \beta \sum_{l=1}^k\lambda_lg_l)^{\frac{1}{\beta}}\,d\mu}
{\int(1+ \beta \sum_{l=1}^k\lambda_lg_l)^{\frac{1}{\beta}}\,d\mu}
\end{equation}
for all $A \in \mathcal{F}$ satisfies the constraints (\ref{eqn:constr_1})
and (\ref{eqn:constr_2}). Furthermore,
the complementary slackness conditions
$$\lambda_i (c_i -\int g_i\,d\nu)=0,$$
hold for $i=1, \ldots, k_1$.
\end{assumption}


\begin{thm}\label{fmom}
Under Assumption \ref{assm2}, $\nu^1$ is
an optimal solution to ${\bf O_2}(\beta)$.
\end{thm}

\begin{remark}
{\em In many cases the inequality (\ref{ineq:200}) can be equivalently expressed as a finite number  of linear constraints on $\lambda_i$'s.
For example, if each $g_i$ has $\mathbb{R}^+$ as its range then clearly (\ref{ineq:200}) is equivalent to $\lambda_i\geq 0$ for all $i$.
As another illustration,
consider $g_i=(x-K_i)^+$ for $i=1,2,3$ and $K_1 \leq K_2 \leq K_3$.
Then (\ref{ineq:200}) is equivalent to
\begin{eqnarray*}
 1+ \beta \lambda_1(K_2-K_1)&\geq& 0,\\
 1+\beta \lambda_1(K_3-K_1) + \beta \lambda_1(K_3-K_2) &\geq& 0,\\
\lambda_1+\lambda_2+\lambda_3 &\geq& 0.\\
\end{eqnarray*}

Note that if each $g_i$ has $(m+1)$-th moment finite,
then (\ref{ineq:201}) holds for $\beta=\frac{1}{m}$.}
\end{remark}

\vspace{0.1in}

{\bf \noindent Existence and Uniqueness of Lagrange multipliers:}
To obtain the optimal distribution by formula (\ref{form2}), we must solve the set of $k$ nonlinear equations given by:
\begin{equation}
\label{eqn:331}
\frac{\int g_j\left(1+\beta \sum_{l=1}^k\lambda_lg_l\right)^{\frac{1}{\beta}}\,d\mu}
{\int\left(1+ \beta \sum_{l=1}^k\lambda_lg_l\right)^{\frac{1}{\beta}}\,d\mu}
=c_j\,\,\,\text{for}\,\,j=1,2,\ldots,k
\end{equation}
for $\lambda_1,\lambda_2,\ldots,\lambda_k$.
In view of Assumption \ref{assm2} we define the set
$$\Lambda:=
\left\{(\lambda_1,\lambda_2,\ldots,\lambda_k)\in\mathbb{R}^k : (\ref{ineq:200})\,\,\text{and}\,\,(\ref{ineq:201}) \,\,\text{hold}.\right\}.$$
A solution $\bmath{\lambda}:=(\lambda_1,\lambda_2,\ldots,\lambda_k)$ to (\ref{eqn:331}) is called \textit{feasible} if lies in the set $\Lambda$.
A feasible solution $\bmath{\lambda}$ is called \textit{strongly feasible} if it further satisfies
\begin{equation}
\label{strong_feasibility}
1+\beta \sum_{l=1}^k\lambda_lc_l>0.
\end{equation}
Theorem~\ref{unik:poly} states sufficient conditions under which a strongly feasible solution to (\ref{eqn:331})
is unique.
\begin{thm}
\label{unik:poly}
Suppose that the variance-covariance matrix of $g_1,g_2,\ldots,g_k$ under any measure $\nu\in\mathcal{P}(\mu)$ is positive definite,
or equivalently, that for any $\nu\in\mathcal{P}(\mu),\sum_{i=1}^k a_ig_i=c\,\text{a.e.}\,(\nu)$, for some constants
$c$ and $a_1,a_2,\ldots,a_k$ implies $c=0$ and $a_i=0$ for all $i=1,2,\ldots,k$.
Then, if a strongly feasible solution to (\ref{eqn:331}) exists, it is unique.
\end{thm}

To solve (\ref{eqn:331}) for $\bmath{\lambda}\in\Lambda$ one may resort to numerical root-search techniques.
A wide variety of these techniques are available in practice (see, for example, \cite{Luenberger2} and \cite{nr}).
In our numerical experiments, we used FindRoot routine from Mathematica which employs a variant of secant method.

Alternatively,
one can use the dual approach of minimizing a convex function over the set $\Lambda$.
In the proof of Theorem~\ref{unik:poly} we have shown one convex function whose
stationary point satisfies a set of equations which is equivalent and related to (\ref{eqn:331}).
Therefore,
it is possible recover a strongly feasible solution to (\ref{eqn:331}) if one exists.
Note that,
contrary to the case with $I$-divergence where the dual convex optimization problem is unconstrained,
here we have a minimization problem of a convex function subject to the constraints (\ref{ineq:200}) and (\ref{strong_feasibility}).

\subsection{Single Random Variable, Single Constraint Setting}

Proposition~\ref{prop:101} below shows the existence of a solution to ${\bf O_2}(\beta)$ under a single random variable, single constraint settings. We then apply this to a few specific examples.

Note that for any random variable $X$,
a non-negative function $g$ and a positive integer $n$,
we always have $E[g(X)^{n+1}] \geq E[g(X)]E[g(X)^{n}]$,
since random variables $g(X)$ and $g(X)^{n}$ are positively associated and hence have non-negative covariance.

\begin{prop} \label{prop:101}
Consider a random variable $X$ with pdf $f$, and a function $g:\mathbb{R}^+\rightarrow \mathbb{R}^+$
such that  $E[g(X)^{n+1}] < \infty$ for a positive integer $n$.
Further suppose that $E[g(X)^{n+1}] > E[g(X)]E[g(X)^{n}]$.
Then the optimization problem:
$$\min_{\tilde{f} \in {{\cal P}(f)}}\int_0^\infty \left(\frac{\tilde{f}(x)}{f(x)}\right)^{1+1/n}f(x)\,dx$$
subject to:
$$\frac{\tilde{E}[g(X)]}{E[g(X)]} = a$$
has a unique solution for  $a\in\left (1,\frac{E[g(X)^{n+1}]}{E[g(X)]E[g(X)^n]}\right)$, given by
$$\tilde{f}(x)=\frac{\left(1+\frac{\lambda}{n}g(x)\right)^nf(x)}{\sum_{k=0}^n n^{-k}\binom{n}{k}E[g(X)^k]\lambda^k}\,\,,x\geq 0,$$
where $\lambda$ is a positive root of the polynomial
\begin{equation}\label{eqn:800}
\sum_{k=0}^n n^{-k}\binom{n}{k}\left\{E[g(X)^{k+1}]-aE[g(X)]E[g(X)^{k}]\right\}\lambda^k=0.
\end{equation}
\end{prop}

As we note in the proof in the Appendix,
the uniqueness  of the solution  follows from Theorem~\ref{unik:poly}.
Standard methods like Newton's, secant or bisection method can be applied to numerically solve (\ref{eqn:800}) for $\lambda$.

\begin{example}
\label{lognormal:exm}
\em{
Suppose $X$ is log-normally distributed with parameters $(\mu,\sigma^2)$
(that is, $\log X$ has normal distribution with mean $\mu$ and variance $\sigma^2$). Then, its density function is
$$f(x)=\frac{1}{x\sqrt{2\pi\sigma^2}}\exp\{-\frac{(\log x-\mu)^2}{2\sigma^2}\}\,\,\,\text{for}\,\,x\geq 0.$$
For the constraint
\begin{equation} \label{eq:sing_const}
\frac{\tilde{E}(X)}{E(X)}=a,
\end{equation}
first consider the case $\beta =\frac{1}{n}=1$. Then,
the probability distribution minimizing the polynomial-divergence is given by:
$$\tilde{f}(x)=\frac{(1+\lambda x)f(x)}{1+\lambda E(X)} \,\,\,\,x\geq 0.$$
From the constraint equation we have:
$$a=\frac{\tilde{E}(X)}{E(X)}=\frac{1}{E(X)+\lambda E(X)^2}\int_0^\infty x(1+\lambda x)f(x)dx=\frac{E(X)+\lambda E(X^2)}{E(X)+\lambda E(X)^2},$$
or
$$\lambda=\frac{E(X)(a-1)}{E(X^2)-aE(X)^2}=\frac{a-1}{e^{\mu+\sigma^2/2}(e^{\sigma^2}-a)}.$$
Since $\frac{E(X)+\lambda E(X^2)}{E(X)+\lambda E(X)^2}$ increases with $\lambda$   and
converges to $\frac{E(X^2)}{E(X)^2}=e^{\sigma^2}$ as $\lambda \rightarrow \infty$ , it follows that our  optimization problem  has a solution if
$a\in[1,e^{\sigma^2})$. Thus if $a>e^{\sigma^2}$ and $\beta=1$, Assumption 2 cannot hold.
Further, it is easily checked that $\frac{E(X^{n+1})}{E(X)E(X^n)}=e^{n\sigma^2}$.
Therefore, for $\beta=\frac{1}{n}$, a solution always exists for $a\in[1,e^{n\sigma^2})$.
}
\end{example}

\begin{example}
\label{gamma:exm}
\em{
Suppose that rv $X$ has a Gamma distribution with density function
$$f(x)=\frac{\theta^{\alpha}x^{\alpha-1} e^{-\theta x}}{\Gamma(\alpha)}\,\,\,\text{for}\,\,x\geq 0$$
and as before,  the constraint is given by (\ref{eq:sing_const}).
Then, it is easily seen that $\frac{E(X^{n+1})}{E(X)E(X^n)}=1+\frac{n}{\alpha}$, so that  a solution with $\beta=\frac{1}{n}$ exists for $a\in[1,1+\frac{n}{\alpha})$.
}
\end{example}

\begin{example}
\label{pareto:exm}
\em{
Suppose $X$ has a Pareto distribution with probability density function:
$$f(x)=\frac{\alpha-1}{(x+1)^\alpha}\,\,\,\text{for}\,\,x\geq 0$$
and as before,  the constraint is given by (\ref{eq:sing_const}).
Then, it is easily seen that
$$\frac{E[X^{n+1}]}{E[X]E[X^n]}=\frac{(\alpha-n-1)(\alpha-2)}{(\alpha-n-2)(\alpha-1)}.$$

As in previous  examples, we see that a probability distribution minimizing the polynomial-divergence with $\beta=\frac{1}{n}$ with $n<\alpha-2$ exists when
$a\in\left[1,\frac{(\alpha-n-1)(\alpha-2)}{(\alpha-n-2)(\alpha-1)}\right)$.
}
\end{example}

\subsection{Weighted Least Squares Approach to find Perturbed Solutions}
\label{wls}

Note that an optimal solution to ${\bf O_2}(\beta)$ may not exist when Assumption~\ref{assm2} does not hold.
For instance, in Proposition~\ref{prop:101}, it may not exist for $a > \frac{E[g(X)^{n+1}]}{E[g(X)]E[g(X)^n]}$.
In that case, selecting $\tilde{f}$ and associated  $\lambda$ so that $\frac{\tilde{E}[g(X)]}{E[g(X)]}$ is very close to
$\frac{E[g(X)^{n+1}]}{E[g(X)]E[g(X)^n]}$ is a reasonable practical strategy.

\vspace{0.1in}

We now discuss how such solutions  may be achieved for general problems using
a weighted least squares approach.
Let $\bmath{\lambda}$ denote $(\lambda_1,\lambda_2,\ldots,\lambda_k)$ and  $\nu_{\bmath{\lambda}}$ denote  the measure defined by right hand side of (\ref{form2}).
We write  ${\bf c}$ for $(c_1,c_2,\ldots,c_k)$
and re-express the optimization problem ${\bf O_2}(\beta)$ as ${\bf O_2}(\beta,{\bf c})$ to explicitly show its dependence on ${\bf c}$.
Define
$$\mathcal{C}:=
\left\{{\bf c}\in\mathbb{R}^k : \exists \,\, \bmath{\lambda}\in\Lambda
\,\,\text{with}\,\,\int g_j\,d\nu_{\bmath{\lambda}}=c_j
\right\}.$$
Consider  ${\bf {c} }\notin\mathcal{C}$.
Then, ${\bf O_2}(\beta,{\bf c})$ has no solution.
In that case, from a practical viewpoint, it is reasonable to consider as a solution
a measure $\nu_{\bmath{\lambda}}$ corresponding to a ${\bf c}'\in\mathcal{C}$ such that
this ${\bf c}'$ is in some sense closest to ${\bf c}$ amongst all elements in $\mathcal{C}$.
To concretize the notion of closeness between two points we define the metric
$$d ({\bf x},{\bf y})=\sum_{i=1}^k\frac{\left(x_i-y_i\right)^2}{w_i} $$
between any two points ${\bf x}=(x_1,x_2,\ldots,x_k)$ and ${\bf y}=(y_1,y_2,\ldots,y_k)$ in $\mathbb{R}^k$,
where ${\bf w}=(w_1,w_2,\ldots,w_k)$ is a constant vector of weights expressing relative importance of the constraints : $w_i>0$ and $\sum_{i=1}^k w_i=1$.

Let ${\bf c_0}:=\arg\inf_{{\bf c}'\in\bar{\mathcal{C}}} d({\bf c},{\bf c}')$, where $\bar{\mathcal{C}}$ is the closure of $\mathcal{C}$. Except in the simplest settings, $\mathcal{C}$ is difficult to explicitly evaluate and so
determining ${\bf c_0}$ is also no-trivial.

The optimization problem below, call it ${\bf \tilde{O} _2}(\beta,t,{\bf c})$,
gives solutions $({{\bf \lambda}_t}, {\bf y}_t)$ such that the vector  ${\bf c}_t$ defined as
$${\bf c}_t:=\left(\int g_1\,d\nu_{\lambda_t},\int g_2\,d\nu_{\lambda_t},\ldots,\int g_k\,d\nu_{\lambda_t}\right),$$
has $d$-distance from ${\bf c}$ arbitrarily close to  $d({\bf c_0},{\bf c})$ when $t$ is sufficiently close to zero.
From implementation viewpoint, the measure $\nu_{\lambda_t}$ may serve as a reasonable surrogate for the optimal solution sought for  ${\bf O_2}(\beta,{\bf c})$.
(Avelleneda et al. \cite{avel3} implement a related least squares strategy to arrive at an updated probability
measure in discrete settings).
\begin{equation}
\min_{\bmath{\lambda}\in\Lambda, {\bf y}} \left[\int\left(\frac{d\nu_{\bmath{\lambda}}}{d\mu}\right)^{\beta+1}\,d\mu +\frac{1}{t}\sum_{i=1}^k\frac{y_i^2}{w_i}\right]
\end{equation}
subject to
\begin{equation}
\label{eqn:3001}
\int g_j\,d\nu_{\bmath{\lambda}} = c_j+y_j\,\,\,\text{for}\,\,j=1,2,\ldots,k.
\end{equation}

This optimization problem penalizes each deviation $y_j$ from $c_j$
by adding appropriately weighted squared deviation term to the objective function.
Let $({{\bf \lambda}_t}, {\bf y}_t)$ denote a solution to
${\bf \tilde{O} _2}(\beta,t,{\bf c})$. This can be seen to exist under a mild condition
that the optimal ${\bf \lambda}$ takes values within a compact set  for any $t$.
Note that then ${\bf c} + {\bf y}_t \in \mathcal{C}$.

\begin{prop} \label{prop:later}
For any ${\bf c} \in \mathbb{R}^k$, the solutions $({{\bf \lambda}_t}, {\bf y}_t)$ to ${\bf \tilde{O} _2}(\beta,t,{\bf c})$ satisfy the relation
$$\lim_{t\downarrow 0}d({\bf c},{\bf c}+{\bf y}_{t}) =  d({\bf c},{\bf c_0}).$$
\end{prop}

\section{Incorporating Constraints on Marginal Distributions}

Next we state and prove the analogue of Theorem 1 and 2 when there is a constraint on a marginal distribution of a few components of the given random vector.
Later in Remark \ref{change:var} we discuss how this generalizes to the case where the constraints involve moments and marginals of functions of the given random vector.
Let $\bold X$ and $\bold Y$ be two random vectors having a law $\mu$ which is given by a joint probability density function\ $f(\bold x, \bold y)$.
Recall that $\mathcal{P}(\mu)$ is the set of all probability measures which are absolutely continuous w.r.t. $\mu$.
If $\nu\in\mathcal{P}(\mu)$ then $\nu$ is also specified by a probability density function say, $\tilde{f}(\cdot)$, such that
$\tilde{f}(\bold x,\bold y)=0$ whenever ${f}(\bold x,\bold y)=0$ and
 $\frac{d\nu}{d\mu}=\frac{\tilde{f}}{f}$.
In view of this we may formulate our optimization problem in terms of probability density functions instead of measures.
Let $\mathcal{P}(f)$ denote the collection of density functions that are absolutely continuous with respect to the density $f$.

\subsection{Incorporating Views on Marginal Using $I$-Divergence}
Formally, our optimization problem ${\bf O_3}$ is:
$$\min_{\nu \in \mathcal{P}(\mu)}\int log\left(\frac{d\nu}{d\mu}\right)\,d\nu=\min_{\tilde{f} \in \mathcal{P}(f)} \int \log \left(\frac{\tilde{f}(\bold x,\bold y)}{f(\bold x,\bold y)}\right) \tilde{f}(\bold x,\bold y) d \bold x d \bold y,$$
subject to:
\begin{equation} \label{eqn:constr_3}
\int_{\bold  y} \tilde{f}(\bold x,\bold y) d \bold y = g(\bold x)\,\,\,\text{for all}\,\,\,\bold x,
\end{equation}
where $g(\bold x)$ is a given marginal density function of $\bold X$, and
\begin{equation} \label{eqn:constr_4}
\int_{\bold x,\bold y} h_i(\bold x, \bold y) \tilde{f}(\bold x,\bold y) d \bold x d \bold y = c_i
\end{equation}
for $i=1,2\ldots,k$.
For presentation convenience, in the remaining paper we only consider equality constraints on moments of functions (as in (\ref{eqn:constr_4})),
ignoring the inequality constraints.
The latter constraints can be easily handled as in Assumptions (1) and (2) by introducing suitable non-negativity and complementary slackness conditions.

Some notation is needed to proceed further.
Let $\bmath{\lambda}= (\lambda_1,\lambda_2,\ldots,\lambda_k)$ and
$$f_{\bmath{\lambda}}(\bold y|\bold x):=
\frac{\exp (\sum_i\lambda_i h_i(\bold x,\bold y)){f}(\bold y|\bold x)}{\int_{\bold y}
 \exp (\sum_j\lambda_i h_i(\bold x,\bold y)) {f}(\bold y|\bold x)d \bold y}=\frac{\exp (\sum_i\lambda_i h_i(\bold x,\bold y)){f}(\bold x,\bold y)}{\int_{\bold y} \exp (\sum_i\lambda_i h_i(\bold x,\bold y)) {f}(\bold x,\bold y)d \bold y}\,.$$
Further, let $f_{\bmath{\lambda}}(\bold x,\bold y)$ denote the joint density function of $(\bold X, \bold Y)$,
$f_{\bmath{\lambda}}(\bold y|\bold x)\times g(\bold x)$ for all $\bold x,\bold y$,
and $E_{\bmath{\lambda}}$ denote the expectation under $f_{\bmath{\lambda}}$.

For a mathematical claim that depends on x, say $S(x)$,
we write $S(x)$ for almost all $x\,\, w.r.t.\,\, g(x)dx$ to mean that  $m_g(x|\,S(x)\,\, \text{is false})=0$,
where $m_g$ is the measure induced by the density $g$.
That is,
$m_g(A)=\int_{A}g(x)\,dx$\,\, for all  measurable subsets $A$.

\begin{assumption}

\label{assm3}

There exists $\bmath{\lambda}\in\mathbb{R}^k$ such that
$$\int_{\bold y} \exp (\sum_i\lambda_i h_i(x,y)) {f}(\bold x,\bold y)d \bold y<\infty$$
for almost all $\bold x$\ w.r.t\ $g(\bold x)d \bold x$\ and the probability density function $f_{\bmath{\lambda}} $ satisfies the constraints given by (\ref{eqn:constr_4}). That is, for all $i=1,2,...,k$, we have
\begin{equation}
\label{eqn:constr_300}
E_{\bmath{\lambda}}[h_i(\bold X,\bold Y)]=c_i\,.
\end{equation}

\end{assumption}


\begin{thm}
\label{exp:marginal}
Under Assumption \ref{assm3}, $f_{\bmath{\lambda}}(\cdot)$ is
an optimal solution to ${\bf O_3}$.
\end{thm}


In Theorem~\ref{thm:exist_unique}, we develop conditions that ensure uniqueness
of a solution to ${\bf O_3}$ once it exists.

\begin{thm} \label{thm:exist_unique}
Suppose that for almost all $\bold x$ w.r.t. $g(\bold x)d \bold x$,
conditional on $\bold X=\bold x$,
no non-zero linear combination of the random variables $h_1(\bold x,\bold Y),h_2(\bold x,\bold Y),\ldots,h_k(\bold x,\bold Y)$
has zero variance w.r.t. the conditional density $f(\bold y|\bold x)$,
or, equivalently, for almost all $\bold x$ w.r.t. $g(\bold x)d \bold x$,
$\sum_i a_ih_i(\bold x,\bold Y)=c$ almost surely ($f(\bold y|\bold x)d \bold y$) for some constants
$c$ and $a_1,a_2,\ldots,a_k$ implies $c=0$ and $a_i=0$ for all $i=1,2,\ldots,k$.
Then, if a solution to the constraint equations (\ref{eqn:constr_300}) exists, it is unique.
\end{thm}

\begin{remark}\label{change:var}  {\em
Theorem \ref{exp:marginal}, as stated, is applicable when the updated marginal distribution of a sub-vector
$\bold X$ of the given random vector $(\bold X,\bold Y)$ is specified.
More generally,  by a routine change of variable technique, similar specification  on a function of the given random vector
can also be incorporated. We now illustrate this.

Let $\bold Z=(Z_1,Z_2,\ldots,Z_N)$ denote a random vector taking values in $S\subseteq\mathbb{R}^N$ and having a (prior) density
function $f_{\bold Z}$. Suppose the constraints are as follows:
\begin{itemize}
\item
$\left(v_1(\bold Z),v_2(\bold Z),\ldots,v_{k_1}(\bold Z)\right)$ have a joint density function given by $g(\cdot)$.
\item
$\tilde{E}[v_{k_1+1}(\bold Z)]=c_1,\,\tilde{E}[v_{k_1+2}(\bold Z)]=c_2,\,\ldots,\,\tilde{E}[v_{k_2}(\bold Z)]=c_{k_2-k_1},$
\end{itemize}
where $0\leq k_1\leq k_2\leq N$ and $v_{1}(\cdot),\,v_{2}(\cdot),\,\ldots,\,v_{k_2}(\cdot)$ are some functions on $S$.

If $k_2< N$ we define $N-k_2$ functions $v_{k_2+1}(\cdot),\,v_{k_2+2}(\cdot),\,\ldots,\,v_{N}(\cdot)$ such that the function
$v:S\rightarrow\mathbb{R}^N$ defined by
$v(\bold z)=\left(v_1(\bold z),v_2(\bold z),\ldots,v_N(\bold z)\right)$ has a nonsingular Jacobian a.e.
That is,
$$J(\bold z):=\text{det}\left(\left(\frac{\partial v_i}{\partial z_j}\right)\right)\neq 0\,\,\text{for almost all}\,\, \bold z\,\,\text{w.r.t.}\,f_{\bold Z},$$
where  we are assuming that the functions $v_{1}(\cdot),\,v_{2}(\cdot),\,\ldots,\,v_{k_2}(\cdot)$ allow such a construction.

Consider
$\bold X = (X_1, \ldots, X_{k_1})$, where $X_i=v_i(\bold Z)$ for $i \leq k_1$ and
$\bold Y = (Y_1, \ldots, Y_{N-k_1})$, where $Y_i=v_{k_1+i}(\bold Z)$ for $i \leq N-k_1.$
Let $f(\cdot, \cdot)$ denote the density function of $(\bold X, \bold Y)$.
Then, by the change of variables formula for densities,

$$f(\bold x,\bold y)=f_{\bold Z}\left(w(\bold x, \bold y)\right)
[J\left(w(\bold x, \bold y)\right)]^{-1}\,,$$

where $w(\cdot)$ denotes the  local inverse function of $v(\cdot)$, that is,
if $v(\bold z) = (\bold x,\bold y)$, then, $\bold z= w(\bold x,\bold y)$.

The constraints can easily be expressed in terms of $(\bold X,\bold Y)$ as\\
$$\bold X \,\,\text{have joint density given by}\,\,g(\cdot)$$
 and
\begin{equation}
\label{eqn:99}
\tilde{E}[Y_i]=c_i \,\, for\,\,i=1,2,\ldots,(k_2-k_1).
\end{equation}

Setting $k=k_2-k_1$, from  Theorem (\ref{exp:marginal}) it follows that  the optimal density
function of $(\bold X,\bold Y)$ as:
$$f_{\bmath{\lambda}}(\bold x,\bold y)=
\frac{e^{\lambda_1y_1+\lambda_2y_2+\cdots+\lambda_ky_k}f(\bold x,\bold y)}
                                {\int_ye^{\lambda_1y_1+\lambda_2y_2+\cdots+\lambda_ky_k}f(\bold x,\bold y)\,d\bold y}\times g(\bold x)\,,$$
where $\bold \lambda_k$'s is chosen to satisfy (\ref{eqn:99}).

Again by the change of variable formula, it follows that the optimal density of $\bold Z$ is given by:
$$\tilde{f}_{\bold Z}(\bold z)=f_{\bmath{\lambda}}(v_1(\bold z),v_2(\bold z),\ldots,v_N(\bold z))J(\bold z)\,\,.\,\,\Box$$

}
\end{remark}

\subsection{Incorporating Constraints on Marginals Using Polynomial-Divergence}

Extending Theorem \ref{exp:marginal} to the case of polynomial-divergence is straightforward. We state the details for completeness.
As in the case of $I$-divergence, the  following notation will simplify our exposition.
Let
\begin{eqnarray*}
f_{\bmath{\lambda},\beta}(\bold y|\bold x):
&=&\frac{\left(1+\beta \left(\frac{f(\bold x)}{g(\bold x)}\right)^\beta \sum_j\lambda_j h_j(\bold x,\bold y)\right)^{\frac{1}{\beta}}{f}(\bold y|\bold x)}{\int_{\bold y} \left(1+\beta \left(\frac{f(\bold x)}{g(\bold x)}\right)^\beta \sum_j\lambda_j h_j(\bold x,\bold y)\right)^{\frac{1}{\beta}}{f}(\bold y|\bold x)d \bold y }\\
&=&\frac{\left(1+\beta\left(\frac{f(\bold x)}{g(\bold x)}\right)^\beta \sum_j\lambda_j h_j(\bold x,\bold y)\right)^{\frac{1}{\beta}}{f}(\bold x,\bold y)}{\int_{\bold y} \left(1+\beta\left(\frac{f(\bold x)}{g(\bold x)}\right)^\beta \sum_j\lambda_j h_j(\bold x,\bold y)\right)^{\frac{1}{\beta}}{f}(\bold x,\bold y)d \bold y }\,.
\end{eqnarray*}
If the marginal of $\bold X$ is given by $g(\bold x)$ then the joint density
$f_{\bmath{\lambda},\beta}(\bold y|\bold x)\times g(\bold x)$ is denoted by $f_{\bmath{\lambda},\beta}(\bold x,\bold y)$.
$E_{\bmath{\lambda},\beta}$ denotes the expectation under $f_{\bmath{\lambda},\beta}(\cdot)$.

Consider  the optimization problem ${\bf O_4}(\beta)$:
$$\min_{\tilde{f} \in {\cal P}(f) } \int \left(\frac{\tilde{f}(\bold x,\bold y)}{f(\bold x,\bold y)}\right)^\beta \tilde{f}(\bold x,\bold y) d \bold x d \bold y\,\,,$$
subject to (\ref{eqn:constr_3}) and (\ref{eqn:constr_4}).

\begin{assumption}

\label{assm4}

There exists $\bmath{\lambda} \in\mathbb{R}^k$ such that
\begin{equation}
\label{ineq:100}
1+\beta\left(\frac{f(\bold x)}{g(\bold x)}\right)^\beta \sum_j\lambda_j h_j(\bold x,\bold y)\geq 0\,\,,
\end{equation}
for almost all $(\bold x,\bold y)$\ w.r.t.\  $f(\bold y|\bold x)\times g(\bold x)d \bold yd \bold x$ and
\begin{equation}
\label{ineq:101}
\int_{\bold y} \left(1+\beta\left(\frac{f(\bold x)}{g(\bold x)}\right)^\beta \sum_j\lambda_j h_j(\bold x,\bold y)\right)^{ 1+\frac{1}{\beta} }{f}(\bold x,\bold y)d \bold y<\infty\,\,,
\end{equation}
for almost all $\bold x$\ w.r.t.\  $g(\bold x)d \bold x$.
Further, the probability density function $f_{\bmath{\lambda},\beta}(\bold x,\bold y)$ satisfies the constraints given by (\ref{eqn:constr_4}). That is, for all $i=1,2,...,k$, we have
\begin{equation}
\label{eqn:constr_301}
E_{\bmath{\lambda},\beta}[h_i(\bold X,\bold Y)]=c_i\,.
\end{equation}
\end{assumption}


\begin{thm} \label{thm:poly_marg}
Under Assumption \ref{assm4}, $f_{\bmath{\lambda},\beta}(\cdot)$ is an optimal solution to ${\bf O_4}(\beta)$.
\end{thm}


Analogous to the discussion  in Remark (\ref{change:var}), by a suitable change of variable, we can adapt the above theorem
to the case where the  constraints involve marginal distribution and/or moments of functions of a given random vector.

We conclude this section with a brief discussion on uniqueness of the solution to ${\bf O_4}(\beta)$.
Any $\bmath{\lambda} \in\mathbb{R}^k$ satisfying (\ref{ineq:100}), (\ref{ineq:101}) and (\ref{eqn:constr_301}) is called a \textit{feasible solution} to ${\bf O_4}(\beta)$.
A feasible solution $\bmath{\lambda}$ is called \textit{strongly feasible} if it further satisfies
$$1+\beta \left(\frac{f(\bold x)}{g(\bold x)}\right)^\beta \sum_j\lambda_j c_j> 0\,\,\,\text{for almost all}\,\,\bold x\,\,\text{w.r.t.}\,\,g(\bold x)d \bold x.$$

The following theorem can be proved using similar arguments as those used to prove Theorem \ref{unik:poly}. We omit the details.
\begin{thm}
Suppose that for almost all $\bold x$ w.r.t. $(g(\bold x)d \bold x)$, conditional on $\bold X= \bold x$, no non-zero linear combination of $h_1(\bold x,\bold Y),h_2(\bold x,\bold Y),\ldots,h_k(\bold x,\bold Y)$ has zero variance under any measures $\nu$ absolutely continuous w.r.t $f(\bold y|\bold x)$.
Or equivalently, that for any measures $\nu$ absolutely continuous w.r.t $f(\bold y|\bold x),\,\, \sum_{i=1}^k a_ih_i(\bold x,\bold Y)=c\,\,$ almost everywhere\,\,$(\nu)$, for some constants
$c$ and $a_1,a_2,\ldots,a_k$, implies $c=0$ and $a_i=0$ for all $i=1,2,\ldots,k$.
Then, if a strongly feasible solution to (\ref{eqn:constr_301}) exists, it is unique.
\end{thm}

Note that when Assumptions 3 and 4 do not hold,
 the weighted least squares methodology developed in Section~3.3 can again be used
 to arrive at a reasonable perturbed solution that may be useful from implementation viewpoint.

\section{Portfolio Modeling  in Markowitz Framework}

 In this section we apply the methodology developed in Section 4.1  to the Markowitz framework:
namely to the setting where there are $N$ assets whose returns under the `prior distribution' are multivariate Gaussian.
Here, we explicitly identify the posterior distribution that incorporates views/constraints on marginal distribution
of some random variables and moment constraints on other random variables.
As mentioned in the introduction, an important application of our approach is that if for a particular portfolio of assets, say an index,
it is established that the return distribution is fat-tailed (specifically, the pdf is a regularly varying function),
say with the density function $g$,
then by using that as a constraint, one can arrive at an updated posterior distribution for all the underlying assets.
Furthermore, we show that if an underlying asset has a non-zero correlation with this portfolio under the prior distribution,
then under the posterior distribution, this asset has a tail distribution similar to that given by $g$.

Let $(\bold X,\bold Y)=(X_1,X_2,\ldots,X_{N-k},Y_1,Y_2,\ldots,Y_k)$ have a $N$ dimensional
multivariate Gaussian distribution with mean
$\bmath{\mu}=(\bmath{\mu}_{\bold x},\bmath{\mu}_{\bold y})$
and the variance-covariance matrix
$$ \bold\Sigma=\left(\begin{array}{cc}\bold\Sigma_{\bold{xx}}&\bold\Sigma_{\bold {xy}}\\ \bold\Sigma_{\bold {yx}}&\bold\Sigma_{\bold{yy}}. \end{array}\right)$$

We consider  a posterior distribution that satisfies the  view that:
$$X\,\,\text{has probability density function}\,\, g(\bold x)\,\,\text{and}\,\,\tilde{E}(\bold Y)=\bold a,$$
where $g(\bold x)$ is a given probability density function on $\mathbb{R}^{N-k}$ with finite first moments
along each component
and $\bold a$ is a given vector in\ $\mathbb{R}^k$.
As we discussed  in Remark~\ref{change:var} (see also Example \ref{6asset:example} in Section 7), when the view is on marginal
distributions of linear combinations of underlying assets, and/or on moments of linear functions of the underlying assets,
the problem can be easily transformed to the above setting by a suitable change of variables.

To find the  distribution of $(\bold X,\bold Y)$ which incorporates the above views, we solve the minimization problem ${\bf O_5}$:

$$\min_{\tilde{f} \in {\cal P}(f)} \int_{(\bold x,\bold y)\in\mathbb{R}^{N-k}\times \mathbb{R}^{k}} \log\left(\frac{\tilde{f}(\bold x,\bold y)}{f(\bold x,\bold y)}\right) \tilde{f}(\bold x,\bold y)\, d\bold xd\bold y$$
subject to the constraint:
$$\int_{\bold y\in\mathbb{R}^{k}} \tilde{f}(\bold x,\bold y) d\bold y=g(\bold x) \,\,\,\, \forall \bold x$$
and
\begin{equation}\label{mom:y}
\int_{\bold x\in\mathbb{R}^{N-k}}\int_{\bold y\in\mathbb{R}^{k}}\bold y \tilde{f}(\bold x,\bold y) d\bold y d\bold x=\bold a,
\end{equation}
where\ $f(\bold x,\bold y)$ is the density of $N$-variate normal distribution denoted by $\mathcal{N}_N(\bmath{\mu},\bold\Sigma)$.


\begin{prop}
\label{normal:marg}
Under the assumption that $\Sigma_{\bold {xx}}$ is invertible, the optimal solution to ${\bf O_5}$ is given by
\begin{equation}\label{post:marg}
\tilde{f}(\bold x,\bold y)=g(\bold x)\times\tilde{f}(\bold y|\bold x)
\end{equation}
where $\tilde{f}(\bold y|\bold x)$ is the probability density function of

$$\mathcal{N}_{k}\left(\bold a+\bold\Sigma_{\bold {yx}}\bold\Sigma_{\bold {xx}}^{-1}(\bold x-E_g[\bold X])\,,\,
      \bold\Sigma_{\bold{yy}}-\bold\Sigma_{\bold {yx}}\bold\Sigma_{\bold {xx}}^{-1}\bold\Sigma_{\bold {xy}}\right)$$
where $E_g(\bold X)$ is the expectation of $X$ under the density function $g$.
\end{prop}


{\bf Tail behavior of the marginals of the posterior distribution:}
We now specialize to the case where $\bold X$ (also denoted by $X$) is a single random variable so that $N=k+1$,
and Assumption~\ref{ass:ass_u_me} below is satisfied by pdf $g$.  Specifically, $(X,\bold Y)$ is distributed as
$\mathcal{N}_{k+1}(\bmath{\mu},\bold\Sigma)$
with
$$\bmath{\mu}^T=(\mu_x,\bmath{\mu}_{\bold y}^T) \,\,\, \text{and}\,\,\, \bold\Sigma=\left(\begin{array}{cc}\sigma_{xx}&\bmath{\sigma}_{x\bold y}^T\\ \bmath{\sigma}_{x\bold y}&\bold\Sigma_{\bold{yy}} \end{array}\right)$$
where\ $\bmath{\sigma}_{x\bold y}=(\sigma_{xy_1},\sigma_{xy_2},...,\sigma_{xy_k})^T$\ with $\sigma_{xy_i}=Cov(X,Y_i).$

\begin{assumption}  \label{ass:ass_u_me}
The pdf $g(\cdot)$ is regularly varying, that is, there exists a constant $\alpha>1$
($\alpha >1$ is needed for $g$ to be integrable) such that
\[
\lim_{t \rightarrow \infty}\frac{g(\eta t)}{g(t)}= \frac{1}{\eta^{\alpha}}
\]
for all $\eta>0$ (see, for instance, \cite{feller:two}). In addition, for any $a\in\mathbb{R}$ and $b\in\mathbb{R}^+$
\begin{equation} \label{eqn:1101}
\frac{g(b(t-s-a))}{g(t)} \leq h(s)
\end{equation}
for some non-negative function $h(\cdot)$ independent of $t$ (but possibly depending on $a$ and $b$) with the property
that $Eh(Z) <\infty$ whenever $Z$ has a Gaussian distribution.
\end{assumption}

\begin{remark}{ \em
Assumption~\ref{ass:ass_u_me} holds, for instance, when $g$ corresponds to $t$-distribution with $n$ degrees of freedom, that is,
$$g(s)=\frac{\Gamma(\frac{n+1}{2})}{\sqrt{n\pi}\Gamma(\frac{n}{2})}(1+\frac{s^2}{n})^{-(\frac{n+1}{2})}\,,$$
Clearly, $g$ is regularly varying with $\alpha = n+1$.
To see (\ref{eqn:1101}), note that
$$\frac{g(b(t-s-a))}{g(t)}= \frac{(1+t^2/n)^{(n+1)/2}}{(1+b^2(t-s-a)^2/n)^{(n+1)/2}}\,\,\,.$$
Putting $t'=\frac{bt}{\sqrt{n}}, s'=\frac{b(s+a)}{\sqrt{n}}$ and $c=\frac{1}{b}$
we have
$$\frac{(1+t^2/n)}{(1+b^2(t-s-a)^2/n)}=\frac{1+c^2t'^2}{1+(t'-s')^2}\,\,\,.$$
Now (\ref{eqn:1101}) readily follows from the fact that
$$\frac{1+c^2t'^2}{1+(t'-s')^2}\leq max\{1,c^2\}+c^2s'^2+c^2|s'|,$$
for any two real numbers $s'$ and $t'$.
To verify the last inequality, note that if $t'\leq s'$ then $\frac{1+c^2t'^2}{1+(t'-s')^2}\leq 1+c^2s'^2$ and if $t'>s'$ then
\begin{eqnarray*}
\frac{1+c^2t'^2}{1+(t'-s')^2}=\frac{1+c^2(t'-s'+s')^2}{1+(t'-s')^2} &=& \frac{1+c^2(t'-s')^2}{1+(t'-s')^2}+c^2s'^2+c^2s'\frac{2(t'-s')}{1+(t'-s')^2}\\
                                                                   &\leq&  max\{1,c^2\}+c^2s'^2+c^2|s'|.
\end{eqnarray*}
Note that if $h(x)=x^{m}$ or $h(x)= \exp(\lambda x)$ for any $m$ or $\lambda$ then the
last condition in Assumption~\ref{ass:ass_u_me} holds.
}
\end{remark}

From Proposition (\ref{normal:marg}), we note  that the posterior distribution of $(X,\bold Y)$ is
$$\tilde{f}(x,\bold y)=g(x)\times\tilde{f}(\bold y|x)$$
where $\tilde{f}(\bold y|x)$ is the probability density function of
$$\mathcal{N}_k\left(\bold a+\left(\frac{x-E_g(X)}{\sigma_{xx}}\right)\bmath{\sigma}_{x\bold y},
\bold\Sigma_{\bold{yy}}-\frac{1}{\sigma_{xx}}\bmath{\sigma}_{x\bold y}\bmath{\sigma}_{x\bold y}^t\right),$$
where $E_g(X)$ is the expectation of $X$ under the density function $g$.
Let $\tilde{f}_{Y_1}$ denote  the marginal density of $Y_1$ under the above posterior distribution.
Theorem~\ref{tail} states  a key result of this section.
\begin{thm}
\label{tail}
Under Assumption~\ref{ass:ass_u_me}, if
$\sigma_{xy_1}\neq 0$,
then
\begin{equation}\label{limit:marg}
\lim_{s\to\infty}\frac{\tilde{f}_{Y_1}(s)}{g(s)}=\left (\frac{\sigma_{xy_1}}{\sigma_{xx}} \right )^{\alpha-1}.
\end{equation}
\end{thm}
 Note that (\ref{limit:marg}) implies that
 $$\lim_{x \rightarrow \infty}\frac{\int_x \tilde{f}_{Y_1}(s) ds} {\int_x g(s) ds} = \left (\frac{\sigma_{xy_1}}{\sigma_{xx}} \right )^{\alpha-1}.$$

\section{Comparing Different Objectives}

Given that in many examples one can use $I$-divergence as well as polynomial-divergence as an objective function for arriving at an updated probability measure, it is natural to compare the optimal solutions in these cases. Note that the total variation distance between two probability measures $\mu$ and $\nu$  defined on $(\Omega, {\cal F})$ equals
\[
\sup \{ \mu(A) - \nu(A)| A \in  {\cal F} \}.
\]
This may also serve as an objective function in our search for a reasonable
probability measure that incorporates expert views and is close to the original
probability measure. This has an added advantage of being a metric (e.g., it satisfies
the triangular inequality).

We now compare these three different types of objectives to get a qualitative flavor
of the differences in the  optimal solutions in two simple settings (a rigorous analysis in general settings may be
a subject for future research).   The first corresponds to the case of   single random variable whose prior distribution  is exponential.
In the second setting,  the views correspond to probability
assignments to  mutually exclusive and exhaustive set of events.

\subsection{Exponential Prior Distribution}

Suppose that the random variable $X$ is exponentially distributed with rate $\alpha$ under $\mu$.
Then its pdf equals
$$f(x)=\alpha e^{-\alpha x},\,x\geq 0.$$

Now suppose that our view is that under the updated measure $\nu$ with density function $\tilde{f}$,
$\tilde{E}(X)=\int x\tilde{f}(x)\,dx=\frac{1}{\gamma} > \frac{1}{\alpha}$.

\vspace{0.1in}

{\bf $I$-divergence:}
When the objective function is to minimize $I$-divergence, the optimal solution
is obtained as an exponentially twisted distribution that satisfies the desired constraint.
It is easy to see that exponentially twisting an exponential distribution with rate
$\alpha$ by an amount
$\theta$ leads to
another exponential distribution with rate $\alpha - \theta$ (assuming that $\theta < \alpha$).
Therefore, in our case
\[
\tilde{f}(x) =\gamma e^{-\gamma x},\,x\geq 0.
\]
satisfies the given constraint and is the solution to this problem.
Note here that the tail distribution function equals
$\exp(-\gamma x)$ and is heavier than $\exp(-\alpha x)$, the original tail distribution of  $X$.

\vspace{0.1in}

 {\bf Polynomial-divergence:}
Now consider the case where the objective corresponds to a polynomial-divergence with parameter equal to $\beta$, i.e, it equals
$$\int \left(\frac{\tilde{f}(x)}{f(x)} \right)^{\beta+1}\,f(x) dx.$$
Under this objective, the optimal pdf is
$$\tilde{f}(x)=\frac{(1+\beta\lambda x)^{1/\beta}\alpha e^{-\alpha x}}
{\int(1+\beta\lambda x)^{1/\beta} \alpha e^{-\alpha x}\,dx},$$
where $\lambda>0$ is chosen so that the mean under
$\tilde{f}$ equals $\frac{1}{\gamma}$.

While this may not have a closed form solution,
it is clear that on a logarithmic scale,
 $\tilde{f}(x)$ is asymptotically similar to
 $\exp(-\alpha x)$ as $x \rightarrow \infty$ and hence has a lighter tail than the solution under the $I$-divergence.

\vspace{0.1in}

 {\bf Total variation distance:}
Under total variation distance as an objective,
 we show that given any $\varepsilon$, we can find a new
density function $\tilde{f}$ so that the mean under the new distribution equals
$\frac{1}{\gamma}$ while the total variation distance is less than $\varepsilon$. Thus the optimal value
of the objective function is zero, although there may be no pdf that attains this value.

To see this, consider,

 $$\tilde{f}(x)=\frac{\varepsilon}{2}\times\frac{I_{(a-\delta,a+\delta)}}{2\delta}+\left(1-\frac{\varepsilon}{2}\right)\alpha e^{-\alpha x}\,\,\,\text{for}\,\,x\geq 0.$$
Then,
\[
\tilde{E}(X)=\int x\tilde{f}(x)\,dx=\left(\frac{\varepsilon}{2}\right)a+\frac{1-\frac{\varepsilon}{2}}{\alpha}.
\]
Thus, given any $\varepsilon$, if we select
$$a=\frac{\frac{1}{\gamma}-\frac{1}{\alpha}}{\frac{\varepsilon}{2}}+\frac{1}{\alpha},$$
we see that
$$\tilde{E}(X)=\left(\frac{\varepsilon}{2}\right)a+\frac{1-\frac{\varepsilon}{2}}{\alpha}=\frac{1}{\gamma}.$$
We now show that total variation distance between $f$ and $\tilde{f}$ is less than $\varepsilon$.
To see this, note that
$$\left|\int_A f(x) dx-\int_A\tilde{f}(x)dx\right|\leq\left(\frac{\varepsilon}{2}\right) P(A)$$
for any set $A$ disjoint from $(a-\delta,a+\delta)$,
where the probability $P$ corresponds to the density $f$.
Furthermore, letting $L(S)$  denote the Lebesgue measure of set $S$,
$$\left|\int_A f(x) dx-\int_A\tilde{f}(x)dx\right| \leq\left(\frac{\varepsilon}{4\delta}\right)L(A)+\left(\frac{\varepsilon}{2}\right) P(A)$$
for any set $A\subset(a-\delta,a+\delta)$.
Thus, for any set $A\subset (0,\infty)$
$$\left|\int_A f(x) dx-\int_A\tilde{f}(x)dx\right|\leq\left(\frac{\varepsilon}{4\delta}\right)L\left(A\cap(a-\delta,a+\delta)\right)+\left(\frac{\varepsilon}{2}\right) P(A).
$$
Therefore,
$$\sup_A \left|\int_A f(x) dx-\int_A\tilde{f}(x)dx\right|\leq\left(\frac{\varepsilon}{2}\right) P(A)+\left(\frac{\varepsilon}{4\delta}\right)2\delta<\varepsilon.$$

This also illustrates that it may be difficult to have an elegant characterization
of solutions under the total variation distance, making the other two as more attractive measures from this viewpoint.

\subsection{Views on Probability of Disjoint Sets} \label{sec:disjoint}

Here, we consider the case where the views correspond to  probability
assignments under posterior measure $\nu$ to  mutually exclusive and exhaustive set of events and
note that objective functions associated with $I$-divergence, polynomial-divergence and total variation distance give identical results.

Suppose that our views correspond to:
$$\nu(B_i)=\alpha_i,\ i=1,2,...k\ \text{where}$$
$$B_i's\ \text{are disjoint},\ \cup B_i=\Omega\ \text{and}\ \sum_{i=1}^k\alpha_i=1.$$

For instance, if $L$ is a continuous random variable denoting loss amount from a portfolio and there
 is a view that value-at-risk at a certain amount $x$ equals $1\%$. This may be modeled
as $\nu\{L \geq x\} = 1\%$ and $\nu\{L < x\} = 99\%$.

\vspace{0.1in}

{\bf $I$-divergence:}
Then, under the $I$-divergence setting, for any event $A$, the optimal
\[
\nu(A)=\frac{ \int_A e^{\sum_i\lambda_i I(B_i)}\,d\mu}
{\int e^{\sum_i\lambda_i I(B_i)}\,d\mu}
=\frac{\sum_i e^{\lambda_i}\mu(A \cap B_i)}{\sum_i e^{\lambda_i}\mu(B_i)}.
\]
Select $\lambda_i$ so that  $e^{\lambda_i}= \alpha_i/\mu(B_i)$.
Then  it follows
that the specified views hold and
\begin{equation} \label{eqn:disjoint}
\nu(A) = \sum_i \alpha_i \mu(A \cap B_i)/\mu(B_i).
\end{equation}

\vspace{0.1in}

{\bf Polynomial-divergence:}
The analysis remains identical when we use polynomial-divergence with parameter $\beta$.
Here, we see that optimal
$$\nu(A)=\frac{\sum_i(1+\beta\lambda_i)^{1/\beta}\mu(A\cap B_i)}{\sum_i(1+\beta\lambda_i)^{1/\beta}\mu(B_i)}.$$
Again, by setting $(1+\beta\lambda_i)^{1/\beta}= \alpha_i/\mu(B_i)$,
(\ref{eqn:disjoint}) holds.

\vspace{0.1in}

{\bf Total variation distance:}
If  the objective is the
total variation distance, then clearly,
the objective function is never less than $\max_i| \mu(B_i) -\alpha_i|$.
We now show that $\nu$  defined by (\ref{eqn:disjoint}) achieves this lower bound.

To see this, note that

\begin{eqnarray*}
|\nu(A)-\mu(A)| & = &
\left|\sum_i \left (\nu(A\cap B_i)-\mu(A \cap B_i) \right )\right| \\
& \leq &  \sum_i | \nu(A\cap B_i)-\mu(A \cap B_i)| \\
& \leq & \sum_i\frac{\mu(A \cap B_i)}{\mu(B_i)}|\alpha_i- \mu(B_i)| \\
& \leq & \max_i| \mu(B_i) -\alpha_i|.
\end{eqnarray*}

\section{Numerical Experiments}
\label{numeric:exp}
Three simple  experiments are conducted.
In the first, we consider a calibration problem,
where the distribution of the underlying Black-Scholes model is updated through polynomial-divergence based optimization to match the observed options prices.
We then consider a portfolio modeling problem in Markowitz framework,
where VAR (value-at-risk) of a portfolio consisting of 6 global indices is evaluated.
Here, the  model parameters  are estimated from historical data.
We then use the proposed methodology to incorporate a view that return from one of the index has a $t$ distribution,
along with views on the moments of returns of some  linear combinations of the indices.
In the third example, we empirically observe the parameter space where Assumption 2 holds,
in a simple two random variable, two constraint setting.

\begin{example}
\em{
Consider a security whose  current price $S_0$ equals $50$. Its volatility $\sigma$ is estimated to equal $0.2$.
Suppose that the   interest rate $r$ is constant and equals $5\%$.
Consider two liquid European call options on this security with  maturity $T=1$ year, strike prices  $K_1=55,\,\text{and}\, K_2=60$, and  market prices of $5.00$ and $3.00$, respectively.
It is easily checked that the Black Scholes price of these options at $\sigma= 0.2$ equals $3.02$ and $1.62$, respectively.
It can also be easily checked that there is no value of $\sigma$  making two of the Black-Scholes prices match the observed market prices.

We apply polynomial-divergence methodology to arrive at a probability measure closest
to the Black-Scholes measure while matching the observed market prices of the two liquid options.
Note that under Black-Scholes $$S(T)\sim\,\,\text{log-normal}\left(\log S(0)+(r-\frac{\sigma^2}{2})T,\sigma^2T\right)=\,\text{log-normal}\left(\log 50 + 0.03, 0.04\right)$$
which is heavy-tailed in that the moment generating function does not exist in the neighborhood of the origin.
Let $f$ denote the pdf for the above log-normal distribution.
We apply Theorem~2 with $\beta=1$  to obtain the posterior distribution:
$$\tilde{f}(x)=\frac{\left(1+\lambda_1(x-K_1)^+ +\lambda_2(x-K_2)^+\right)f(x)}{\int_0^\infty\left(1+\lambda_1(x-K_1)^+ +\lambda_2(x-K_2)^+\right)f(x)\,dx}$$
where  $\lambda_1$ and $\lambda_2$ are solved from the constrained equations:
\begin{equation}
\label{eqn:666}
\tilde{E}[e^{-rT}(S(T)-K_1)^+]=5.00\,\,\,\,\text{and}\,\,\,\, \tilde{E}[e^{-rT}(S(T)-K_2)^+]=3.00.
\end{equation}

The solution comes out to be  $\lambda_1=0.0945945$ and $\lambda_2=-0.0357495$ (found using FindRoot of Mathematica).
Note that $\lambda_1>-\frac{1}{K_2-K_1}=-0.2$ and $\lambda_1+\lambda_2>0$.
Therefore these values are all feasible.
Furthermore, since $\lambda_1\times 5+\lambda_2\times 3>0$,
they are strongly feasible as well.
Plugging these  values  in $\tilde{f}(\cdot)$ we get the posterior density that can be used to price other options of the same maturity.
The second row of Table~\ref{tab:price} shows the resulting European call option prices for different values of strike prices under this posterior distribution.

\begin{table}[ht]
\begin{center}
\begin{tabular}{|c|c|c|c|c|c|c|c|}
\hline
Strike&50&55&60&65&70&75&80 \\
\hline
BS&5.2253&3.0200&1.62374&0.8198&0.3925&0.1798&0.0795 \\
\hline
Posterior I&7.6978&5.0000&3.0000&1.6779&0.8851&0.4443&0.2139\\
\hline
Posterior II&8.0016&4.9698&3.0752&1.9447&1.15306&0.63341&0.3276\\
\hline
Posterior III&8.0000&4.9991&3.0201&1.8530&1.0757&0.5821&0.2977\\
\hline
Posterior IV&7.7854&4.8243&3.0584&1.9954&1.2092&0.6743&0.3525\\
\hline

\end{tabular}
\caption{\label{tab:price} Option prices for different strikes as computed by the Black Scholes and different posterior distributions of the form given by (\ref{form2}).
Here, BS stands for Black Scholes price at $\sigma=0.2$.
Posterior I is the optimal distribution corresponding to the two constraints (\ref{eqn:666}).
Posterior II (resp. III,and IV)  is the posterior distribution obtained by solving the perturbed problem with equal weights
(resp., increasing weights and decreasing weights) given to the four constraints given by (\ref{eqn:666}) and (\ref{eqn:777}).
}
\end{center}
\end{table}
}

Now suppose that the market prices of two more European options of same maturity with strike prices  50 and 65
are found to be  8.00 and 2.00, respectively.
In the above  posterior distribution these prices equal  7.6978 and 1.6779, respectively.
To arrive at a density function 
that agrees with these prices as well,  we solve the associated four constraint problem by adding
 the following constraint equations to (\ref{eqn:666}):
\begin{equation}
\label{eqn:777}
\tilde{E}[e^{-rT}(S(T)-K_0)^+]=8.00\,\,\,\,\text{and}\,\,\,\, \tilde{E}[e^{-rT}(S(T)-K_3)^+]=2.00,
\end{equation}
where $K_0=50$ and $K_3=65$.

With these added constraints, we observe that the optimization problem ${\bf O_2}(\beta)$ lacks any solution of the form (\ref{form2}).
We then implement the proposed weighted least squares methodology to arrive at a perturbed solution.
With weights $w_1=w_2=w_3=w_4=\frac{1}{4}$ and $t=4\times 10^{-3}$ (so that each $tw_i$ equals $10^{-3}$),
the new posterior is of the form (\ref{form2}) with $\lambda$-values given by
$\lambda_1=0.334604,\lambda_2=-0.445519,\lambda_3=-0.0890854$ and $\lambda_4=0.409171$.
The third row of Table~\ref{tab:price} gives the option prices under this new posterior.
The last two rows report the option prices under the posterior measure resulting from weight combinations
$t{\bf w}=(10^{-6},10^{-5},10^{-4},10^{-3})$ and $t{\bf w}=(10^{-3},10^{-4},10^{-5},10^{-6})$, respectively.
\end{example}

\begin{example}
\label{6asset:example}
\em{
We consider an equally weighted portfolio in six global indices:
ASX (the S$\&$P/ASX200, Australian stock index),
DAX (the German stock index),
EEM (the MSCI emerging market index),
FTSE (the FTSE100, London Stock Exchange),
Nikkei (the Nikkei225, Tokyo Stock Exchange)
and
S$\&$P (the Standard and Poor 500).
Let $Z_1,Z_2,\ldots,Z_6$ denote the weekly rate of returns\footnote{Using logarithmic rate of return gives almost identical results.}
from ASX, DAX, EEM, FTSE, Nikkei and S$\&$P, respectively.
We take prior distribution of $(Z_1,Z_2,\ldots,Z_6)$ to be multivariate Gaussian with mean vector and variance-covariance matrix estimated from historical
prices of these indices for the last two years (161 weeks, to be precise) obtained from Yahoo-finance.
Assuming a notional amount of 1 million, the {value-at-risk} (VaR) for our portfolio for  different confidence levels is shown in the second column of Table \ref{var}.

Next, suppose our view is that DAX will have an expected weekly return of $0.2\%$ and will have a $t$-distribution with $3$ degrees of freedom.
Further suppose that we expect all the indices to strengthen and give  expected  weekly rates of return as in Table \ref{view_mean}.
For example, the third row in that table expresses the view that the rate of return from emerging market will be higher than that of FTSE by $0.2\%$.
Expressed mathematically:
$$\tilde{E}[Z_3 - Z_4]=0.002,$$
where $\tilde{E}$ is the expectation under the posterior probability.
The other rows may be similarly interpreted. 

We define new variables as $X=Z_2$, $Y_1=Z_1$, $Y_2=Z_3-Z_4$, $Y_3=Z_4$, $Y_4=Z_5$ and $Y_5=Z_6$.
Then our views are $\tilde{E}[Y]=(0.1\%,0.2\%,0.1\%,0.2\%,0.1\%)^t$
and $\frac{X-0.002}{\sqrt{Var(X)}}$ has a standard $t$-distribution with 3 degrees of freedom.

The third column in Table \ref{var} reports  VaRs at different confidence levels under the posterior distribution incorporating the only views on the expected returns
(i.e, without the view that $X$ has a $t$-distribution).
We see that these do not differ much  from those under the prior distribution.
This is because the views on the expected returns have little effect on the tail:
the posterior distribution is Gaussian and even though the mean has shifted (variance remains the same) the tail probability remains negligible. Contrast this with the effect of incorporating the view that $X$ has a $t$-distribution.
The VaRs (computed from $100,000$ samples) under this posterior distribution are reported in the last column.


\begin{table}[t]
\begin{center}
\begin{tabular}{|c|c|}
\hline
Index& average rate of return \\
\hline
ASX& 0.001\\
\hline
DAX& 0.002\\
\hline
EMM - FTSE& 0.002\\
\hline
FTSE& 0.001\\
\hline
Nikkei& 0.002\\
\hline
S$\&$P& 0.001\\
\hline
\end{tabular}
\caption{\label{view_mean} Expert view on average weekly return for different indices.}
\end{center}
\end{table}

\begin{table}[t]
\begin{center}
\begin{tabular}{|c|c|c|c|}
\hline
VaR at & Prior & Posterior 1 & Posterior 2 \\
\hline
$99\%$& 8500 & 8400 & 16000\\
\hline
$97.5\%$& 7200 & 7100 &  11200\\
\hline
$95\%$& 6000 & 5900 & 8200\\
\hline
\end{tabular}
\caption{\label{var}  The second column reports VaR under the prior distribution.
The third  reports VaR under the posterior distribution incorporating views on expected returns only.
The last column reports VaR under the posterior distribution incorporating views on expected returns as well as
a view that returns from DAX are  $t$-distributed with three degrees of freedom.}
\end{center}
\end{table}
}

\end{example}

\begin{example}
\label{existance:example}
\em{
In this example we further refine the observation made in Proposition 2 and the following examples that typically
the solution space where  Assumption~2 holds increases
with increasing $n=\frac{1}{\beta}$. We note that even in simple cases, this need not always be true.

Specifically, consider random variables  $X$ and $Y$  such that

$$\left[\begin{array}{c}\log X\\\log Y\end{array}\right]\sim \text{bivariate Gaussian}\left(\left[\begin{array}{c}0\\0\end{array}\right],\left[\begin{array}{cc}1&\rho\\\rho&1\end{array}\right]\right).$$

Then, $X$ and $Y$ are log-normally distributed and their
joint density function of $(X,Y)$ is given by:
$$\frac{1}{2\pi xy\sqrt{1-\rho^2}} {\exp}\left[-\frac{1}{2(1-\rho^2)}\{(\log x)^2+(\log y)^2-2\rho(\log x)(\log y)\}\right].$$
Consider the constraints
$$\left[\begin{array}{c}\tilde{E}(X)\\\tilde{E}(Y)\end{array}\right]=\left[\begin{array}{c}aE(X)\\bE(Y)\end{array}\right].$$
Our goal is to find values of $a$ and $b$ for which the associated optimization problem ${\bf O_2(\beta)}$ has a solution.
The probability distribution minimizing the polynomial-divergence with $\beta=\frac{1}{n}$ is of the form:
$$\tilde{f}(x,y)=\frac{(1+\frac{\lambda}{n}x+ \frac{\xi}{n}y)^n f(x,y)}{\int_0^\infty (1+\frac{\lambda}{n}x+ \frac{\xi}{n}y)^n f(x,y)\,dxdy}
=\frac{(1+\frac{\lambda}{n}x+ \frac{\xi}{n}y)^n f(x,y)}{E[(1+\frac{\lambda}{n}X+ \frac{\xi}{n}Y)^n]}.$$
Now from the constraint equations we have
$$a=\frac{E[X(1+\frac{\lambda}{n}X+ \frac{\xi}{n}Y)^n]}{E[X]E[(1+\frac{\lambda}{n}X+ \frac{\xi}{n}Y)^n]}$$
and
$$b=\frac{E[Y(1+\frac{\lambda}{n}X+ \frac{\xi}{n}Y)^n]}{E[Y]E[(1+\frac{\lambda}{n}X+ \frac{\xi}{n}Y)^n]}.$$
Note that since $X$ and $Y$ takes values in $(0,\infty)$ only $\lambda\geq 0$ and $\xi\geq 0$ are feasible.
Using ParametricPlot of Mathematica we plot the values of
$$\left(\frac{E[X(1+\frac{\lambda}{n}X+ \frac{\xi}{n}Y)^n]}{E[X]E[(1+\frac{\lambda}{n}X+ \frac{\xi}{n}Y)^n]},\frac{E[Y(1+\frac{\lambda}{n}X+ \frac{\xi}{n}Y)^n]}{E[Y]E[(1+\frac{\lambda}{n}X+ \frac{\xi}{n}Y)^n]}\right)$$
for $\lambda$ and $\xi$ in the range $[0,10 n]$. Figure (\ref{fig:range-corr}), depicts the range when
$\rho=-\frac{1}{4},\,\rho=0,\,\rho=\frac{1}{4}\,\text{and}\,\rho=\frac{1}{2}$ respectively, for $n=1,n=2\,\text{and}\,n=3.$

\begin{figure}[ht]
\begin{center}
\subfigure[$\rho=-\frac{1}{4}$]{
\includegraphics[width=5cm]{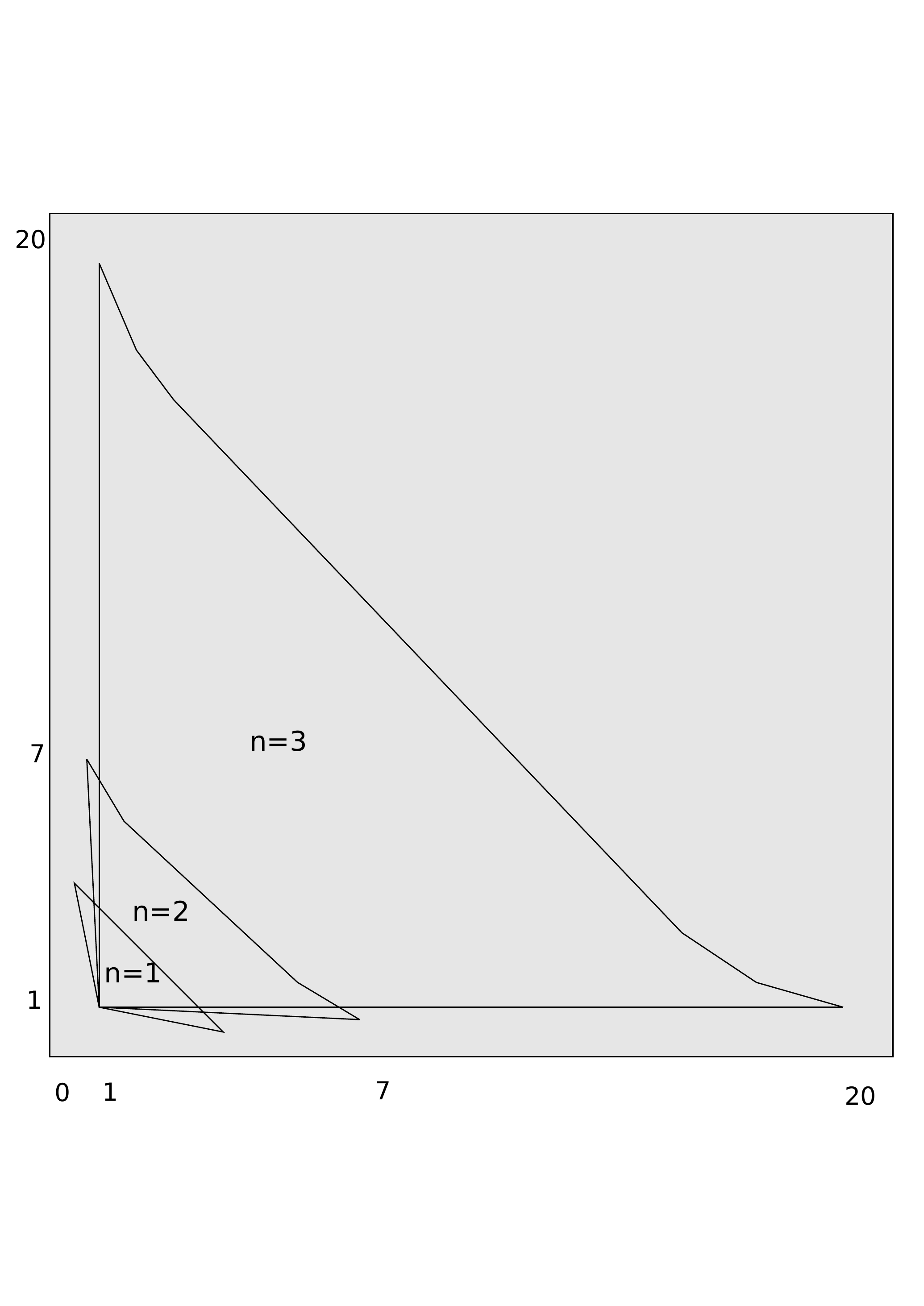}}
\subfigure[$\rho=0$]{
\includegraphics[width=5cm]{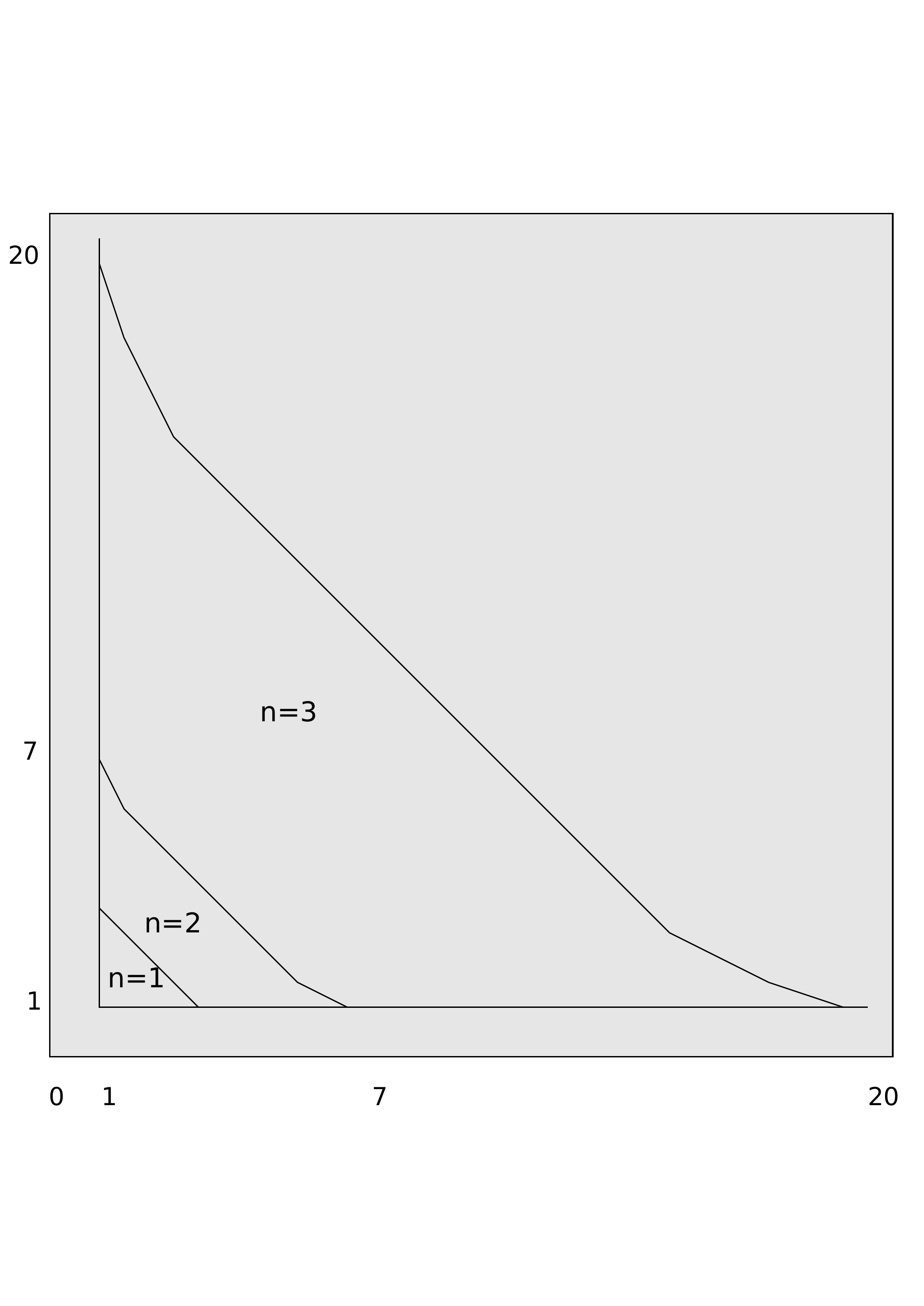}}
\subfigure[$\rho=\frac{1}{4}$]{
\includegraphics[width=5cm]{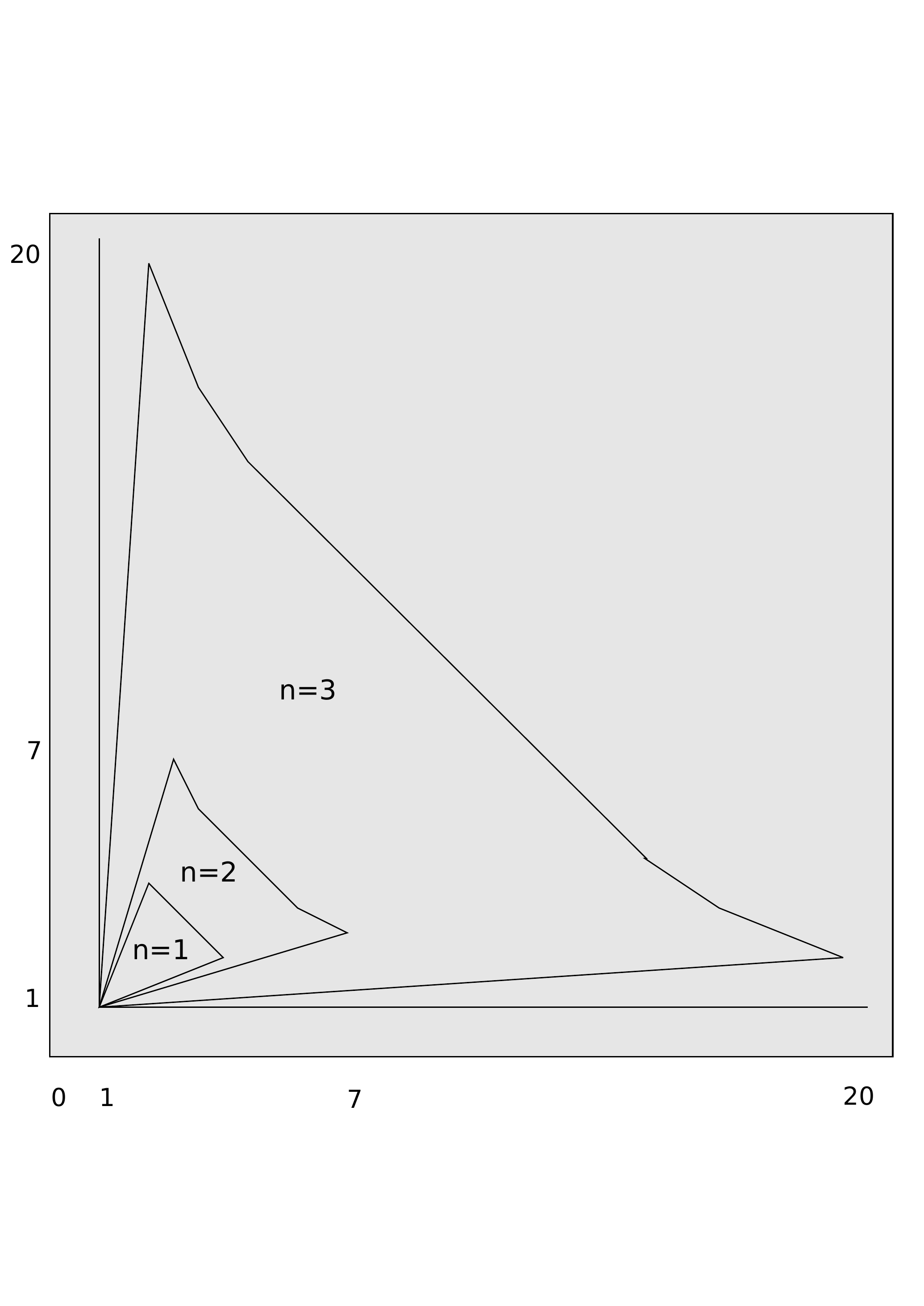}}
\subfigure[$\rho=\frac{1}{2}$]{
\includegraphics[width=5cm]{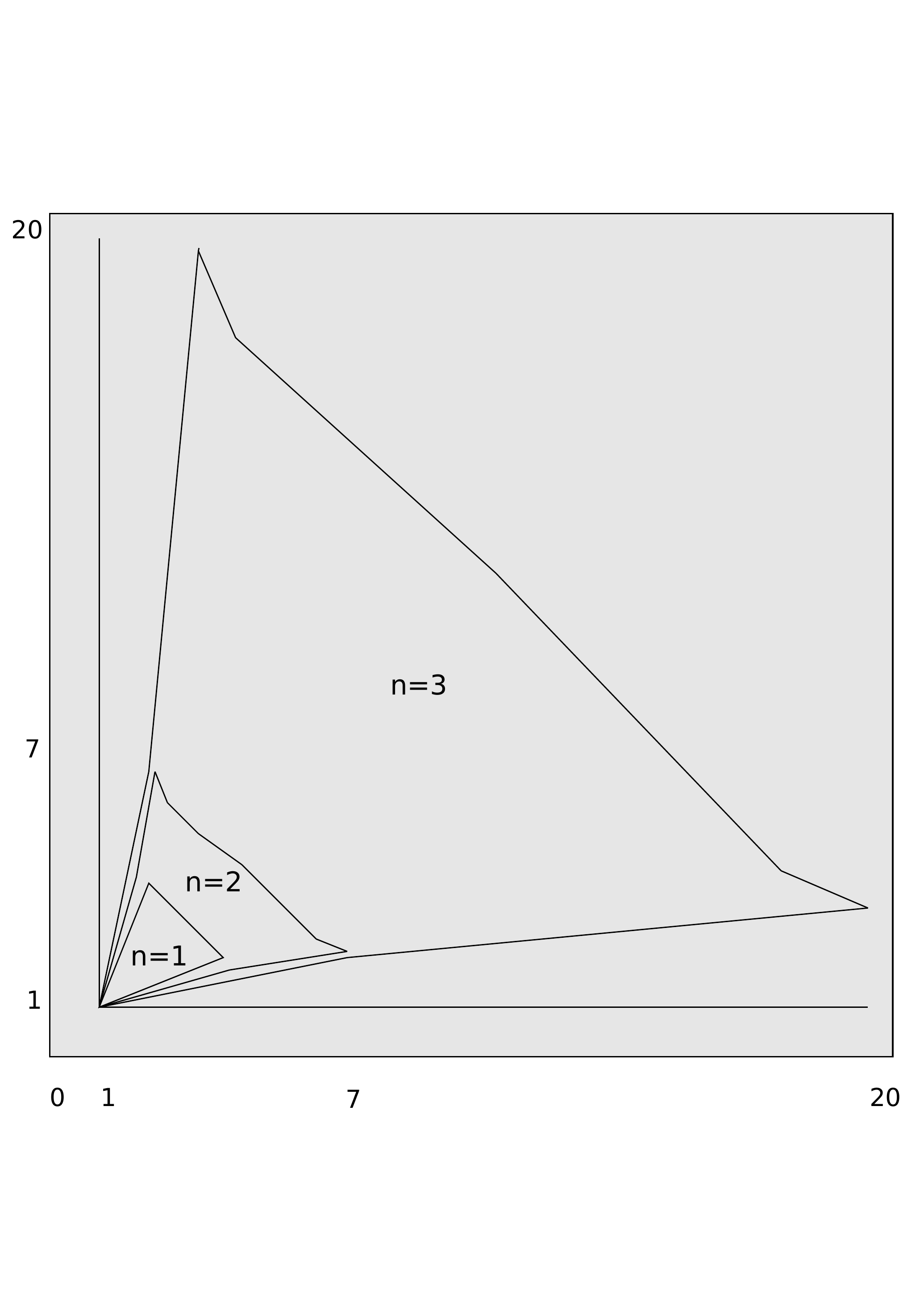}}
\caption{\label{fig:range-corr}Solution range as a function of correlation and $n$.}
\end{center}
\end{figure}

From the graph it appears that the solution space strictly increases with $n$ when $\rho \geq 0$. However, this is not true for $\rho < 0$.

}
\end{example}

\section{Conclusion}
In this article,
we built upon existing methodologies that use relative entropy based ideas
for incorporating mathematically specified views/constraints to a given financial model to arrive at a more accurate one,
when the underlying random variables are light tailed.
In the existing financial literature,
these ideas have found many applications  including in portfolio modeling, in model calibration and in ascertaining the pricing probability measure in incomplete markets.
Our key contribution was to show that under technical conditions,
using polynomial-divergence,
such constraints may be uniquely incorporated even when the underlying random variables have fat tails.
We also extended the proposed methodology to allow for constraints on marginal distributions of functions of underlying variables.
This, in addition to the constraints on moments of functions of underlying random variables,
traditionally considered for such problems.
Here, we considered,  both $I$-divergence and polynomial-divergence as objective.

We also specialized our results to the Markowitz portfolio modeling framework
where multivariate Gaussian distribution is used to model asset returns.
Here, we developed close form solutions for the updated posterior distribution.
In case when there is a constraint that a marginal of a single portfolio of assets has a fat-tailed distribution,
we showed that under the posterior distribution,
marginal of all assets with non-zero correlation with this portfolio have similar fat-tailed distribution.
This may be a reasonable and a simple way to incorporate realistic tail behavior in a portfolio of assets.

We also qualitatively compared the solution to the optimization problem in a simple setting
of exponentially distributed prior when the objective function was $I$-divergence,
polynomial-divergence and total variational distance.
We found that in certain settings,
$I$-divergence may put more mass in tails compared to polynomial-divergence,
which may penalize tail deviation more.
Finally, we numerically tested the proposed methodology on some simple examples.

\vspace{0.2in}

{\noindent {\bf Acknowledgement:}
\em The authors would like to thank Paul Glasserman for many directional suggestions as well as specific inputs on the manuscript that greatly helped this effort.
We would also like to thank the Associate Editor and the referees for feedbacks that substantially improved the article.
}

\section{Appendix: Proofs}

\noindent {\bf Proof of Proposition~\ref{prop:001}}
Recall that $f(x)=\frac{\alpha-1}{(1+x)^\alpha}$ for $x\geq 0$, and
$${f}_{\lambda_M}(x)=\frac{e^{\lambda_M x}f(x) I_{[0,M]}(x)}{\int_0^M e^{\lambda_M x}f(x)dx}$$
where $\lambda_M$ is the unique solution to $\int_0^M x {f}_{\lambda}(x)\,dx=c$.

We first show that $\lambda_M \rightarrow 0$ as $M \rightarrow \infty$. To this end, let
$$g(\lambda,M)=\frac{\int_0^M x e^{\lambda x}f(x)\,dx}{\int_0^M e^{\lambda x}f(x)\,dx}\,\,,M\geq 1,\,\lambda\geq 0.$$

We have
\begin{eqnarray*}
\frac{\partial g}{\partial\lambda}
&=&\frac{\left(\int_0^M x^2 e^{\lambda x}f(x)\,dx\right)\left(\int_0^M e^{\lambda x}f(x)\,dx\right)-\left(\int_0^M x e^{\lambda x}f(x)\,dx\right)^2}
{\left(\int_0^M e^{\lambda x}f(x)\,dx\right)^2}\\
&=&\int_0^M x^2 \left(\frac{e^{\lambda x}f(x)}{\int_0^M e^{\lambda x}f(x)\,dx}\right)\,dx-
\left(\int_0^M x \left(\frac{e^{\lambda x}f(x)}{\int_0^M e^{\lambda x}f(x)\,dx}\right)\,dx\right)^2\\
&=&E_{\lambda}[X^2]-\left(E_{\lambda}[X]\right)^2>0
\end{eqnarray*}
where $E_{\lambda}$ is the expectation operator w.r.t density function $f_{\lambda}$.

Also
\begin{eqnarray*}
\frac{\partial g}{\partial M}
&=&\frac{\left(Me^{\lambda M}f(M)\right)\left(\int_0^M e^{\lambda x}f(x)\,dx\right)-
\left(\int_0^M x e^{\lambda x}f(x)\,dx\right)\left(e^{\lambda M}f(M)-f(0)\right)}
{\left(\int_0^M e^{\lambda x}f(x)\,dx\right)^2}\\
&=&\frac{e^{\lambda M}f(M)\int_0^M (M-x)e^{\lambda x}f(x)\,dx+f(0)\int_0^M e^{\lambda x}f(x)\,dx}{\left(\int_0^M e^{\lambda x}f(x)\,dx\right)^2}
>0.
\end{eqnarray*}
Since $\lambda_M$ satisfies $g(\lambda_M,M)=c$, it follows that increasing $M$ leads to reduction in
 $\lambda_M$. That is, $\lambda_M$ is a non-increasing function of $M$.

Suppose that  $\lambda_M \downarrow c_1 >0$.
Then $c=g(\lambda_M,M)\geq g(c_1,M)$ for all $M$.
But since $c_1>0$ we have $g(c_1,M)\rightarrow\infty$ as $M\rightarrow\infty$, a contradiction.
Hence, $\lambda_M\rightarrow 0$ as $M\rightarrow\infty$.

Next, since
$$\int_0^M\log\left(\frac{f_\lambda(x)}{f(x)}\right)f_\lambda(x)\,dx=\lambda\int_0^M xf_\lambda(x)\,dx - \log\left(\int_0^M e^{\lambda x}f(x)\,dx\right),$$
we see that
$$\int_0^M\log\left(\frac{f_{\lambda_M}(x)}{f(x)}\right)f_{\lambda_M}(x)\,dx=\lambda_M c- \log\left(\int_0^M e^{{\lambda_M} x}f(x)\,dx\right).$$
Hence, to prove that the LHS converges to zero as
$M \rightarrow \infty$, it suffices to show
 that\\
$\int_0^M e^{\lambda_M x}f(x)\,dx\rightarrow 1$ or that
$\int_0^M \frac{e^{\lambda_M x}}{(1+x)^{\alpha}}\,dx\rightarrow\frac{1}{\alpha-1}$.

Note that the constraint equation can be re-expressed as:
$$c=\int_0^M xf_{\lambda_M}(x)\,dx=
\frac{\int_0^M\frac{x e^{{\lambda_M} x}}{(1+x)^{\alpha}}\,dx}{\int_0^M\frac{e^{{\lambda_M} x}}{(1+x)^{\alpha}}\,dx}=
\frac{\int_0^M\frac{ e^{{\lambda_M} x}}{(1+x)^{\alpha-1}}\,dx-\int_0^M\frac{ e^{{\lambda_M} x}}{(1+x)^{\alpha}}\,dx}{\int_0^M\frac{e^{{\lambda_M} x}}{(1+x)^{\alpha}}\,dx}=
\frac{\int_0^M\frac{ e^{{\lambda_M} x}}{(1+x)^{\alpha-1}}\,dx}{\int_0^M\frac{e^{{\lambda_M} x}}{(1+x)^{\alpha}}\,dx}-1,$$
or
\begin{equation}
\label{eqn:996}
\frac{\int_0^M\frac{ e^{{\lambda_M} x}}{(1+x)^{\alpha-1}}\,dx}{\int_0^M\frac{e^{{\lambda_M} x}}{(1+x)^{\alpha}}\,dx}=1+c.
\end{equation}

Further, by integration by parts of the numerator, we observe that
\[
{\int_0^M\frac{ e^{{\lambda_M} x}}{(1+x)^{\alpha-1}}\,dx}
= \frac{1}{\lambda_M} \left( \frac{e^{\lambda_M M}}{(1+M)^{\alpha-1}}-1 +(\alpha-1)
\int_0^M\frac{e^{\lambda_M x}}{(1+x)^{\alpha}}\,dx \right).
\]
From the above equation and (\ref{eqn:996}), it follows that
\begin{equation}
\label{eqn:997}
 \int_0^M\frac{e^{\lambda_M x}}{(1+x)^{\alpha}}\,dx=\frac{\frac{e^{\lambda_M M}}{(1+M)^{\alpha-1}}-1}{\lambda_M(c+1)-(\alpha-1)}.
\end{equation}
Since $\lambda_M\rightarrow 0$, it suffices to show  that
$$\frac{e^{\lambda_M M}}{(1+M)^{\alpha-1}}\rightarrow 0.$$
Suppose this is not true. Then,  there exists an  $\eta>0$ and
a sequence $ M_i \uparrow \infty$  such that
$\frac{e^{\lambda_{M_i}M_i}}{(1+M_i)^{\alpha-1}}\geq\eta$.
Equation (\ref{eqn:996}) may be re-expressed as:
\begin{equation}
\label{eqn:998}
\int_0^M\frac{e^{\lambda_M x}}{(1+x)^{\alpha-1}}\left[1-\frac{1+c}{1+x}\right]dx=0
\end{equation}
Given an arbitrary  $K >0$,
one can find an  $M_i \geq 1+2c$ (so that   $1-\frac{1+c}{1+x}\geq \frac{1}{2}$ when $x\geq M_i$) such that, for $x \in[M_i-K, M_i]$
\[
\frac{e^{\lambda_{M_i} x}}{(1+x)^{\alpha-1}}
 \geq \frac{e^{\lambda_{M_i} (M_i-K)}}{(1+M_i)^{\alpha-1}}
\geq \frac{\eta}{2}.
\]
Re-expressing the LHS of (\ref{eqn:998}) evaluated at $M=M_i$ as
\[
\int_0^c \frac{e^{\lambda_{M_i} x}}{(1+x)^{\alpha-1}}\left[1-\frac{1+c}{1+x}\right]dx
+ \int_c^{M_i-K}\frac{e^{\lambda_{M_i} x}}{(1+x)^{\alpha-1}}\left[1-\frac{1+c}{1+x}\right]dx
+ \int_{M_i-K}^{M_i}\frac{e^{\lambda_{M_i} x}}{(1+x)^{\alpha-1}}\left[1-\frac{1+c}{1+x}\right]dx,
\]
we see that this is bounded from below by
\[
-c^2 e^{\lambda_{M_i} c} + K \eta/4.
\]
For sufficiently large $K$, this is greater than zero providing the desired contradiction to (\ref{eqn:998}).$\Box$

\vspace{0.2in}


\textbf{Proof of Theorem~\ref{fmom}:\ \ }
Let $\xi = \int\left(1+ \beta \sum_l\lambda_lg_l\right)^{1/\beta}\,d\mu$.
and $\hat{\lambda}_l = (\beta+1) \beta \lambda_l/\xi^{\beta}$.
Consider  the Lagrangian $\mathcal{L}(\nu)$ for  ${\bf O_2}(\beta)$
defined as
\begin{equation} \label{eqn:700}
\int \left(\frac{d\nu}{d\mu}\right)^{\beta+1}\,d\mu- \sum_l \hat{\lambda}_l\left(\int g_l\,d\nu-c_l\right).
\end{equation}

We first argue that $\mathcal{L}(\nu)$ is a convex function of $\nu$.
Given that $\lambda_l\int g_l\,d\nu$ are linear in $\nu$,
it suffices to show that $\int (\frac{d\nu}{d\mu})^{\beta+1}\,d\mu$
is a convex function of $\nu$.

Note that for $0 \leq s \leq 1$,
\[
\int\left(\frac{d(s\nu_1+(1-s)\nu_2)}{d\mu}\right)^{\beta+1}\,d\mu
\]
equals
\[
\int\left(s\frac{d\nu_1}{d\mu}+(1-s)\frac{d\nu_2}{d\mu}\right)^{\beta+1}\,d\mu
\]
which in turn is dominated by
\[
\int\left[s\left(\frac{d\nu_1}{d\mu}\right)^{\beta+1}+(1-s)\left(\frac{d\nu_2}{d\mu}\right)^{\beta+1}\right]\,d\mu
\]
which equals
\[
s\int\left(\frac{d\nu_1}{d\mu}\right)^{\beta+1}\,d\mu+(1-s)\int\left(\frac{d\nu_2}{d\mu}\right)^{\beta+1}\,d\mu.
\]
Therefore, the Lagrangian $\mathcal{L}(\nu)$ is a convex function of $\nu$.

We now  prove that $\mathcal{L}(\nu)$
is minimized at  $\nu^1$. For this, all we need to show is
that we cannot improve by moving in any feasible direction away from
$\nu^1$. Since, $\nu^1$
satisfies all the constraints, the result then follows.
We now show this.

Let  $f$ denote $\frac{d\nu}{d\mu}$ and  $f^1=\frac{d\nu^1}{d\mu}$.
 Note that (\ref{eqn:700})
 may be re-expressed as
\[
\int \left(f^{\beta+1}- \sum_l\hat{\lambda}_l g_lf\right)\,d\mu + \sum_l \hat{\lambda}_l c_l.
\]

For any  $\nu \in \mathcal{P}(\mu)$ and $t\in[0,1]$
consider the function
\[
G_\nu(t)=\mathcal{L}\left((1-t)\nu^1+t\nu\right).
\]
This in turn equals
\[
\int \left[\{(1-t)f^1+tf\}^{\beta+1}- \sum_l\hat{\lambda}_l g_l\{(1-t)f^1+tf\}\right]\,d\mu
+ \sum_l \hat{\lambda}_l c_l.
\]
We now argue that
${\frac{d}{dt}}_{t=0}G_\nu(t)=0$. Then from this and convexity
of $\mathcal{L}$, the result follows.

To see this, note that ${\frac{d}{dt}}_{t=0}G_\nu(t)$ equals
\begin{equation}
\label{1storder}
\int\left\{(\beta + 1) (f^1)^\beta -\sum_l\hat{\lambda}_l g_l\right\}(f-f^1)\,d\mu.
\end{equation}

 Due to the definition of  $f_1$ and $\hat{\lambda}_i$, it follows that
 the term inside the braces in the integrand in (\ref{1storder}) is a constant.
Since both $\nu^1$ and $\nu$ are probability measures,
therefore ${\frac{d}{dt}}_{t=0}G_\nu(t)=0$ and the result follows.$\Box$

\vspace{0.2in}

\textbf{Proof of Theorem~\ref{unik:poly}:\ \ }
Let $A$ denote the set of all strongly feasible solutions to (\ref{eqn:331}).
Consider the following set of equations:
\begin{equation}
\label{eqn:theta}
\frac{\int g_j\left(1+ \beta \sum_{l=1}^k\theta_l(g_l-c_l)\right)^{\frac{1}{\beta}}\,d\mu}{\int\left(1+
\beta\sum_{l=1}^k\theta_l(g_l-c_l)\right)^{\frac{1}{\beta}}\,d\mu}
=c_j\,\,\,\text{for}\,\,j=1,2,\ldots,k.
\end{equation}

We say that $\bmath{\theta} = (\theta_1,\theta_2,\ldots,\theta_k)$
 is a \textit{strongly feasible} solution to (\ref{eqn:theta}) if it solves  (\ref{eqn:theta}), and
 lies in the set:
$$\left\{\bmath{\theta}\in\mathbb{R}^k\mid 1+\beta \sum_{l=1}^k\theta_l(g_l-\theta_l)\geq0\,\,a.e. \mu,
\int(1+ \beta \sum_{l=1}^k\theta_l(g_l-c_l))^{\frac{1}{\beta}+1}d\mu<\infty\,\,\text{and}\,\beta \sum_{l=1}^k\theta_lc_l<1\right\}.$$
Let $B$ denote the set of all strongly feasible solutions to (\ref{eqn:theta}).

Let $\phi:A\rightarrow B$ and $\psi:B\rightarrow A$ be the mappings defined by
$$\phi(\bmath{\lambda})
=\left(\frac{\lambda_1}{1+\beta \sum_{l=1}^k\lambda_lc_l},\frac{\lambda_2}{1+\beta \sum_{l=1}^k\lambda_lc_l},\ldots,\frac{\lambda_k}{1+ \beta \sum_{l=1}^k\lambda_lc_l}\right)$$
and
$$\psi(\bmath{\theta})
=\left(\frac{\theta_1}{1-\beta \sum_{l=1}^k\theta_lc_l},\frac{\theta_2}{1-\beta \sum_{l=1}^k\theta_lc_l},\ldots,\frac{\theta_k}{1- \beta \sum_{l=1}^k\theta_lc_l}\right).$$
It is easily checked that mapping $\phi$ is a bijection with inverse $\psi$.
Note that if
$\bmath{\lambda} \in A$, then $\phi(\bmath{\lambda})\in B$. To see this, simply divide the numerator and the denominator in (\ref{eqn:331}) by $(1+ \beta \sum_{l=1}^k\lambda_lc_l)^{1/\beta}$.
Conversely, if $\bmath{\theta}\in B$, then $\psi(\bmath{\theta})\in A$. To see this,
 divide the numerator and the denominator in (\ref{eqn:theta}) by $(1- \beta \sum_{l=1}^k\theta_lc_l)^{1/\beta}$.

Therefore,  its suffices to prove uniqueness of strongly feasible solutions to (\ref{eqn:theta}).
To this end, consider the function
 $G:B\rightarrow\mathbb{R}^+$:

$$G(\bmath{\theta})=\int\left(1+ \beta \sum_{l=1}^k\theta_l(g_l-c_l)\right)^{\frac{1}{\beta}+1}\,d\mu.$$
Then,
\begin{equation}
\label{1st:der}
\frac{\partial G}{\partial\theta_i}=\left({1+\beta}\right)\int(g_i-c_i)\left(1+\beta \sum_{l=1}^k\theta_l(g_l-c_l)\right)^{\frac{1}{\beta}}\,d\mu.
\end{equation}
and
$$\frac{\partial^2 G}{\partial\theta_j\partial\theta_i}={\beta}({1+\beta})
\int(g_i-c_i)(g_j-c_j)\left(1+\beta \sum_{l=1}^k\theta_l(g_l-c_l)\right)^{\frac{1}{\beta}-1}\,d\mu\,.$$
We see that the last integral can be written as $E_\theta[(g_i-c_i)(g_j-c_j)]$ times a positive constant independent of $i$ and $j$, where $E_\theta$ denotes expectation under
the measure$$\frac{\left(1+\beta \sum_{l=1}^k\theta_l(g_l-c_l)\right)^{\frac{1}{\beta}-1}\,d\mu}
{\int\left(1+ \beta \sum_{l=1}^k\theta_l(g_l-c_l)\right)^{\frac{1}{\beta}-1}\,d\mu}.$$
Now from the identity
$$E_\theta[(g_i-c_i)(g_j-c_j)]=\text{Cov}_\theta[g_i,g_j]+\left(E_\theta[g_i]-c_i\right)\left(E_\theta[g_j]-c_j\right)$$
and the assumption on $g_i$'s it follows that the Hessian of the function $G$ is positive definite. Thus, $G$ is strictly convex in its domain of definition, that is in $B$. Therefore if a solution to the equation
\begin{equation}
\label{eqn:gradient}
\left(\frac{\partial G}{\partial\theta_1},\frac{\partial G}{\partial\theta_2},\ldots,\frac{\partial G}{\partial\theta_k}\right)=0
\end{equation}
exists in $B$, then it is unique. From (\ref{1st:der}) it follows that the set of equations given by (\ref{eqn:gradient}) and (\ref{eqn:theta}) are equivalent. $\Box$

\vspace{0.2in}

\textbf{ Proof of Proposition~\ref{prop:later}:}
First assume that ${\bf c}\notin\mathcal{C}$ so that ${\bf O_2}(\beta,{\bf c})$  has no solution.
Note that ${\bf c_0}$ may not belong to $\mathcal{C}$ so that ${\bf O_2}(\beta,{\bf c_0})$ may also not have a solution.
Let $B(x,r)$ denote the open ball centered at $x$ and radius $r$ with respect to the metric $d$ defined at Section~\ref{wls}.
For arbitrary $\varepsilon>0$,
let ${\bf c}_\varepsilon\in B({\bf c_0},\varepsilon)\bigcap\mathcal{C}$.
Then ${\bf O_2}(\beta,{\bf c}_\varepsilon)$ has a solution, say $\nu_\varepsilon$ and
let ${\bf \lambda}(\epsilon)$ denote the associated parameters.
It follows that $({\bf \lambda}(\epsilon),{\bf c}_\varepsilon-{\bf c})$ is feasible for
${\bf \tilde{O} _2}(\beta,t,{\bf c})$ for any $t$.
Since $({{\bf \lambda}_t}, {\bf y}_t)$ is a solution to ${\bf \tilde{O} _2}(\beta,t,{\bf c})$,
we have
$$I_\beta(\nu_{\lambda_t}||\mu) + \frac{1}{t}\sum_{i=1}^k\frac{{\bf y}_t(i)^2}{w_i}
\leq
I_\beta(\nu_\varepsilon||\mu) + \frac{1}{t}\sum_{i=1}^k\frac{\left({\bf c}_\varepsilon(i)- {\bf c}(i) \right)^2}{w_i},$$
or
$$I_\beta(\nu_{\lambda_t}||\mu) + \frac{1}{t} d({\bf c},{\bf c}+{\bf y}_t)
\leq
I_\beta(\nu_\varepsilon||\mu) + \frac{1}{t} d({\bf c},{\bf c}_\varepsilon).$$
But,
by triangle inequality and definition of ${\bf c}_\varepsilon$,
it follows that
$$d({\bf c},{\bf c}_\varepsilon)\leq d({\bf c},{\bf c_0}) + d({\bf c_0},{\bf c}_\varepsilon) \leq d({\bf c},{\bf c_0}) + \varepsilon.$$
Therefore,
$$I_\beta(\nu_{\lambda_t}||\mu) + \frac{1}{t} d({\bf c},{\bf c}+{\bf y}_t)
\leq
I_\beta(\nu_\varepsilon||\mu) + \frac{1}{t} d({\bf c},{\bf c_0}) + \frac{\varepsilon}{t}.$$
Since,
$I_\beta(\nu_{\lambda_t}||\mu)\geq 0$ we have
$$d({\bf c},{\bf c}+{\bf y}_t)\leq t I_\beta(\nu_\varepsilon||\mu) + d({\bf c},{\bf c_0}) + \varepsilon.$$
Since $t$ can be chosen arbitrarily small,
we conclude that
\begin{equation}
\label{eqn:421}
d({\bf c},{\bf c}+{\bf y}_t)\leq d({\bf c},{\bf c_0}) + 2\varepsilon.
\end{equation}
Now,
since
${\bf c}+{\bf y}_t\in\mathcal{C}$,
by definition of ${\bf c_0}$ we have
$$d({\bf c},{\bf c}+{\bf y}_t)\geq d({\bf c},{\bf c_0}).$$
Together with inequality (\ref{eqn:421}),
we have,
$$\lim_{t\downarrow 0}d({\bf c},{\bf c}+{\bf y}_t) =  d({\bf c},{\bf c_0}).$$

If ${\bf c}\in\mathcal{C}$ then ${\bf O_2}(\beta,{\bf c})$ has a solution.
Then the above analysis simplifies: ${\bf c_0}={\bf c}$ and each ${\bf c}_\varepsilon$ can be taken to be equal to ${\bf c}$.
We conclude that
$$\lim_{t\downarrow 0}d({\bf c},{\bf c}+{\bf y}_t)=0$$
or ${\bf y}_t\rightarrow 0$.
$\Box$

\vspace{0.2in}

\textbf{Proof of Theorem~\ref{exp:marginal}:\ \ }
In view of (\ref{eqn:constr_3}), we may fix the marginal distribution of $\bold X$ to be $g(\bold x)$ and re-express the objective as

$$\min_{\tilde{f}(\cdot |\bold x) \in {\cal P}({f}(\cdot |\bold x)), \forall \bold x} \int_{\bold x,\bold y} \log\left(\frac{\tilde{f}(\bold y|\bold x)}{f(\bold y|\bold x)}\right)\tilde{f}(\bold y|\bold x) g(\bold x) d \bold y d \bold x+\int_{\bold x} \log\left(\frac{g(\bold x)}{f(\bold x)}\right)g(\bold x)d \bold x\,.$$

The second integral is a constant and can be dropped from the objective. The first integral may in turn be expressed as

$$\int_{\bold x} \min_{\tilde{f}(\cdot |\bold x) \in {\cal P}({f}(\cdot |\bold x))} \left (\int_{\bold y} \log \frac{\tilde{f}(\bold y|\bold x)}{f(\bold y|\bold x)} \tilde{f}(\bold y|\bold x)  d \bold y \right )g(\bold x) d \bold x\,.$$
Similarly the moment constraints can be re-expressed as

$$\int_{\bold x,\bold y} h_i(\bold x,\bold y) \tilde{f}(\bold y|\bold x)g(\bold x) d \bold x d \bold y = c_i,\,\,\,i=1,2,...,k$$
or$$\int_{\bold x}\left (\int_{\bold y} h_i(\bold x,\bold y) \tilde{f}(\bold y|\bold x)d \bold y \right )g(\bold x)d \bold x = c_i,\,\,\,i=1,2,...,k\,.$$
Then, the Lagrangian for this $k$ constraint problem is,

$$\int_{\bold x}
 \left [ \min_{\tilde{f}(\cdot |\bold x) \in {\cal P}({f}(\cdot |\bold x))} \int_{\bold y} \left ( \log \frac{\tilde{f}(\bold y|\bold x)}{f(\bold y|\bold x)} \tilde{f}(\bold y|\bold x)
 -\sum_i\delta_i h_i(\bold x,\bold y)\tilde{f}(\bold y|\bold x) \right )d \bold y  \right ] g(\bold x) d \bold x+\sum_i\delta_i c_i\,.$$

Note that by Theorem (\ref{klmom})
$$\min_{\tilde{f}(\cdot |\bold x) \in {\cal P}({f}(\cdot |\bold x))} \int_{\bold y} \left ( \log \frac{\tilde{f}(\bold y|\bold x)}{f(\bold y|\bold x)} \tilde{f}(\bold y|\bold x)
 -\sum_i\delta_i h_i(\bold x,\bold y)\tilde{f}(\bold y|\bold x) \right )dy$$
has the solution
$$\tilde{f}_{\bmath{\delta}}(\bold y|\bold x)= \frac{\exp (\sum_i\delta_i h_i(\bold x,\bold y)){f}(\bold y|\bold x)}{\int_{\bold y} \exp (\sum_i\delta_i h_i(\bold x,\bold y)) {f}(\bold y|\bold x)d \bold y}=\frac{\exp (\sum_i\delta_i h_i(\bold x,\bold y)){f}(\bold x,\bold y)}{\int_{\bold y} \exp (\sum_i\delta_i h_i(\bold x,\bold y)) {f}(\bold x,\bold y)d \bold y},$$
where we write $\bmath{\delta}$ for $(\delta_1,\delta_2,\ldots,\delta_k)$.
Now taking $\bmath{\delta}=\bmath{\lambda}$, it follows from Assumption \ref{assm3} that
$f_{\bmath{\lambda}}(\bold x,\bold y)=\tilde{f}_{\bmath{\lambda}}(\bold y|\bold x)g(\bold x)$ is a solution to ${\bf O_3}\,.\Box$

\vspace{0.2in}

\textbf{Proof of Theorem~\ref{thm:exist_unique}:\ \ }
Let $F:\mathbb{R}^k\rightarrow \mathbb{R}$ be a function defined as
$$F(\bmath{\lambda})=\int_{\bold x}\log\left(\int_{\bold y} exp\left(\sum_l\lambda_l h_l(\bold x, \bold y)\right)f(\bold y| \bold x)d \bold y\right)g(\bold x)d \bold x-\sum_l\lambda_l c_l.$$
Then,
\begin{eqnarray*}
\frac{\partial F}{\partial\lambda_i}&=&\int_{\bold x}\left(\frac{\int_{\bold y} h_i(\bold x, \bold y) exp\left(\sum_l\lambda_l h_l(\bold x,\bold y)\right)f(\bold y|\bold x)d \bold y}
                                                    {\int_{\bold y} exp\left(\sum_l\lambda_l h_l(\bold x,\bold y)\right)f(\bold y|\bold x)d \bold y}\right)g(\bold x)d \bold x - c_i\\
                                    &=&\int_{\bold x}\left(\int_{\bold y} h_i(\bold x,\bold y) \frac{exp\left(\sum_l\lambda_l h_l(\bold x,\bold y)\right)f(\bold y|\bold x)}
                                                                        {\int_{\bold y} exp\left(\sum_l\lambda_l h_l(\bold x,\bold y)\right)f(\bold y|\bold x)d \bold y}\,d \bold y\right)g(\bold x)d \bold x - c_i\\
                                    &=&\int_{\bold x}\left(\int_{\bold y} h_i(\bold x, \bold y)f_{\bmath{\lambda}}(\bold y|\bold x)d \bold y\right)g(\bold x)d \bold x - c_i\\
                                    &=&\int_{\bold x}\int_{\bold y} h_i(\bold x,\bold y)f_{\bmath{\lambda}}(\bold x, \bold y)d \bold xd \bold y - c_i \\
                                    &=&E_{\bmath{\lambda}}[h_i(\bold X,\bold Y)]-c_i .\\
\end{eqnarray*}
Hence the set of equations given by (\ref{eqn:constr_300}) is equivalent to:
\begin{equation}
\label{eqn:301}
\left(\frac{\partial F}{\partial\lambda_1},\frac{\partial F}{\partial\lambda_2},\ldots,\frac{\partial F}{\partial\lambda_k} \right)=0\,.
\end{equation}

Since
$$\frac{\partial}{\partial\lambda_j}f_{\bmath{\lambda}}(\bold y|\bold x)=h_j(\bold x,\bold y)f_{\bmath{\lambda}}(\bold y|\bold x)-\left(\int_{\bold y} h_j(\bold x,\bold y)f_{\bmath{\lambda}}(\bold y|\bold x)d \bold y\right)
\times f_{\bmath{\lambda}}(\bold y|\bold x),$$
we have
\begin{eqnarray*}
\frac{\partial^2 F}{\partial\lambda_j\partial\lambda_i}&=&\int_{\bold x}\left(\int_{\bold y} h_i(\bold x,\bold y)\frac{\partial}{\partial\lambda_j}f_{\bmath{\lambda}}(\bold y|\bold x)\,d \bold y\right)g(\bold x)\,d \bold x\\
                                                       &=&\int_{\bold x}\left(\int_{\bold y} h_i(\bold x,\bold y)h_j(\bold x,\bold y)f_{\bmath{\lambda}}(\bold y|\bold x)d \bold y\right)g(\bold x)d \bold x\\
                                        &\,& -\int_{\bold x}\left(\int_{\bold y} h_j(\bold x,\bold y)f_{\bmath{\lambda}}(\bold y|\bold x)d\bold y\right)\left(\int_{\bold y} h_i(\bold x,\bold y)f_{\bmath{\lambda}}(\bold y|\bold x)d \bold y\right)g(\bold x)d \bold x\\
                                       &=&E_{g(\bold x)}\left[E_{\bmath{\lambda}}[h_i(\bold X,\bold Y)h_j(\bold X,\bold Y)\mid \bold X]\right]-
                                      E_{g(\bold x)}\left[E_{\bmath{\lambda}}[h_j(\bold X,\bold Y)\mid \bold X]\times E_{\bmath{\lambda}}[h_i(\bold X,\bold Y)\mid \bold X]\right]\\
                                       &=&E_{g(\bold x)}\left[\text{Cov}_{\bmath{\lambda}}[h_i(\bold X,\bold Y),h_j(\bold X,\bold Y)\mid \bold X]\right]\\
\end{eqnarray*}
Where $E_{g(\bold x)}$ denote expectation with respect to the density function $g(\bold x)$. By our assumption, it follows that the Hessian of $F$ is positive definite. Thus,
the function $F$ is strictly convex in $\mathbb{R}^k$.
Therefore if there exist a solution to (\ref{eqn:301}), then it is unique. Since (\ref{eqn:301}) is equivalent to (\ref{eqn:constr_300}), the theorem follows.$\Box$

\vspace{0.2in}

\textbf{Proof of Theorem~\ref{thm:poly_marg}:\ \ }
Fixing the marginal of $\bold X$ to be $g(\bold x)$ we express the objective as
$$\min_{\tilde{f}(\cdot | \bold x) \in {\cal P}(f(\cdot |\bold x)), \forall \bold x} \int_{\bold x,\bold y}\left(\frac{\tilde{f}(\bold y|\bold x)g(\bold x)}{f(\bold y|\bold x)f(\bold x)}\right)^\beta \tilde{f}(\bold y|\bold x) g(\bold x) d\bold y d \bold x\,.$$
This may in turn be expressed as

$$\int_{\bold x} \min_{\tilde{f}(\cdot |\bold x)\in {\cal P}(f( \cdot |\bold x))} \left(\int_{\bold y} \left(\frac{\tilde{f}(\bold y|\bold x)}{f(\bold y|\bold x)}\right)^\beta \tilde{f}(\bold y|\bold x)  d \bold y \right)\left(\frac{g(\bold x)}{f(\bold x)}\right)^\beta g(\bold x) d \bold x\,.$$
Similarly, the moment constraints can be re-expressed as
$$\int_{\bold x}\left (\int_{\bold y} h_i(\bold x,\bold y)\left(\frac{f(\bold x)}{g(\bold x)}\right)^\beta  \tilde{f}(\bold y|\bold x)d \bold y \right )\left(\frac{g( \bold x)}{f(\bold x)}\right)^\beta g(\bold x)d \bold x = c_i,\,\,\,i=1,2,...,k\,.$$

Then, the Lagrangian for this $k$ constraint problem is, up to the constant $\sum_i\delta_i c_i$,

$$\int_{\bold x}
 \left [ \min_{\tilde{f}(\cdot |\bold x)\in {\cal P}(f(\cdot |\bold x))} \int_{\bold y} \left(\left(\frac{\tilde{f}(\bold y|\bold x)}{f(\bold y|\bold x)}\right)^\beta \tilde{f}(\bold y|\bold x)
 -\sum_i\delta_i h_i(\bold x,\bold y)\left(\frac{f(\bold x)}{g(\bold x)}\right)^\beta \tilde{f}(\bold y|\bold x) \right)d \bold y  \right ]\left(\frac{g(\bold x)}{f(\bold x)}\right)^\beta  g(\bold x) d \bold x\,.$$
By Theorem (\ref{fmom}), the inner minimization has the solution $f_{\bmath{\delta},\beta}(\bold y|\bold x)$. Now taking $\bmath{\delta}=\bmath{\lambda}$, it follows from Assumption
(\ref{assm4}) that
$f_{\bmath{\lambda},\beta}(\bold x,\bold y)=f_{\bmath{\lambda}, \beta}(\bold y|\bold x)g(\bold x)$ is the solution to ${\bf O_4}(\beta)$.
$\Box$

\vspace{0.2in}

\textbf{Proof of Proposition~\ref{prop:101}:\ \ }
In view of Assumption (\ref{assm2}), we note that
$1+\frac{\lambda}{n}g(x) \geq 0$ for all $x\geq 0$ if $\lambda\geq 0$.
By Theorem (\ref{fmom}), the probability distribution minimizing the polynomial-divergence (with $\beta=1/n$) w.r.t. $f$ is given by:
$$\tilde{f}(x)=\frac{\left(1+\frac{\lambda}{n}g(x)\right)^nf(x)}{c}\,\,,x\geq 0,$$
where
$$c=\int_0^\infty\left(1+\frac{\lambda}{n}g(x)\right)^nf(x)\,dx=\sum_{k=0}^n n^{-k}\binom{n}{k}E[g(X)^k]\lambda^k.$$
From the constraint equation we have
$$a=\frac{\tilde{E}[g(X)]}{E[g(X)]}=\frac{\int_0^\infty g(x)\left(1+\frac{\lambda}{n}g(x)\right)^nf(x)dx}{cE[g(X)]}
=\frac{\sum_{k=0}^n n^{-k}\binom{n}{k}E[g(X)^{k+1}]\lambda^k}{\sum_{k=0}^n n^{-k}\binom{n}{k}E[g(X)]E[g(X)^k]\lambda^k}.$$

Since, $E[g(X)^{n+1}] > a E[g(X)]E[g(X)^{n}]$, the $n$-th degree term in (\ref{eqn:800}) is strictly positive and the constant term is negative so there exists a positive $\lambda$ that solves this equation.
Uniqueness of the solution now follows from Theorem~\ref{unik:poly}.
$\Box$.

\vspace{0.2in}

\textbf{Proof of Proposition~\ref{normal:marg}:\ \ }
By Theorem~\ref{exp:marginal}:
$$\tilde{f}(\bold x,\bold y)=g(\bold x)\times\tilde{f}(\bold y|\bold x)$$
where
$$\tilde{f}(\bold y|\bold x)=\frac{e^{\bmath{\lambda}^t\bold y}f(\bold y|\bold x)}{\int e^{\bmath{\lambda}^t\bold y}f(\bold y|\bold x)d\bold y}.$$
Here the superscript $t$ corresponds to the transpose.
Now\  $f(\bold y|\bold x)$ is the $k$-variate normal density with mean vector:
$$\bmath{\mu}_{\bold y|\bold x}=\bmath{\mu}_{\bold y}+\bold\Sigma_{\bold {yx}}\bold\Sigma_{\bold {xx}}^{-1}(\bold x-\bmath{\mu}_{\bold x})$$
and the variance-covariance matrix:
$$\bold\Sigma_{\bold y|\bold x}=\bold\Sigma_{\bold{yy}}-\bold\Sigma_{\bold {yx}}\bold\Sigma_{\bold {xx}}^{-1}\bold\Sigma_{\bold {xy}}.$$

Hence\ $\tilde{f}(\bold y|\bold x)$\ is the normal density with mean\ $(\bmath{\mu}_{\bold y|x}+\bold\Sigma_{\bold y|\bold x}\bmath{\lambda})$\ and variance-covariance matrix\ $\bold\Sigma_{\bold y|\bold x}$. Now the moment constraint equation (\ref{mom:y}) implies:

\begin{eqnarray*}
\bold a&=&\int_{\bold x\in\mathbb{R}^{N-k}}\int_{\bold y\in\mathbb{R}^{k}}\bold y \tilde{f}(\bold x,\bold y) d\bold y d\bold x\\
       &=&\int_{\bold x\in\mathbb{R}^{N-k}}g(\bold x)\left(\int_{\bold y\in\mathbb{R}^{k}}\bold y \tilde{f}(\bold y | \bold x) d\bold y\right) d\bold x\\
       &=&\int_{\bold x\in\mathbb{R}^{N-k}}g(\bold x)\left(\bmath{\mu}_{\bold y|\bold x}+\bold\Sigma_{\bold y|\bold x}\bmath{\lambda}\right)d\bold x\\
       &=&\int_{\bold x\in\mathbb{R}^{N-k}}g(\bold x)\left(\bmath{\mu}_{\bold y}+\bold\Sigma_{\bold {yx}}\bold\Sigma_{\bold {xx}}^{-1}(\bold x-\bmath{\mu}_{\bold x})
              +\bold\Sigma_{\bold y|\bold x}\bmath{\lambda}\right)d\bold x\\
       &=&\bmath{\mu}_{\bold y}+\bold\Sigma_{\bold {yx}}\bold\Sigma_{\bold {xx}}^{-1}(E_g[\bold X]-\bmath{\mu}_{\bold x})
                 +\bold\Sigma_{\bold y|\bold x}\bmath{\lambda}.\\
\end{eqnarray*}

Therefore, to satisfy the moment constraint, we must take
$$\bmath{\lambda}=\bold\Sigma_{\bold y|\bold x}^{-1}\left[\bold a-\bmath{\mu}_{\bold y}-
\bold\Sigma_{\bold {yx}}\bold\Sigma_{\bold {xx}}^{-1}(E_g[\bold X]-\bmath{\mu}_{\bold x})\right]\,.$$

Putting the above value of\ $\bmath{\lambda}$\ in\ $(\bmath{\mu}_{\bold y|\bold x}+\bold\Sigma_{\bold y|\bold x}\bmath{\lambda})$\ we see that $\tilde{f}(\bold y|\bold x)$\ is the normal density with mean
$$\bold a+\bold\Sigma_{\bold {yx}}\bold\Sigma_{\bold {xx}}^{-1}(\bold x-E_g[\bold X])$$
and variance-covariance matrix\ $\bold\Sigma_{\bold y|\bold x}.\Box$

\vspace{0.2in}

\textbf{Proof of Theorem~\ref{tail}:\ \ }
We have
$$\tilde{f}(\bold y|x)=D \exp\{-\frac{1}{2}(\bold y-\tilde{\bmath{\mu}}_{\bold y|x})^t\bold\Sigma_{\bold y|x}^{-1}(\bold y-\tilde{\bmath{\mu}}_{\bold y|x})\}$$
for an appropriate constant $D$,
where  $\tilde{\bmath{\mu}}_{\bold y|x}$ denotes $\bold a+\left(\frac{x-E_g(X)}{\sigma_{xx}}\right)\bmath{\sigma}_{x\bold y}$.

Suppose that the stated assumptions hold for $i=1$.
Under the optimal distribution, the marginal density of $Y_1$ is

$$\tilde{f}_{Y_1}(y_1)=\int_{(x,y_2,...,y_k)}D \exp\{-\frac{1}{2}(\bold y-\tilde{\bmath{\mu}}_{\bold y|x})^t\bold\Sigma_{\bold y|x}^{-1}(\bold y-\tilde{\bmath{\mu}}_{\bold y|x})\}g(x)dxdy_2...dy_k.$$

Now the limit in (\ref{limit:marg})  is equal to:

$$\lim_{y_1\to\infty}\int_{(x,y_2,y_3,...,y_k)}D \exp\{-\frac{1}{2}(\bold y-\tilde{\bmath{\mu}}_{\bold y|x})^t\bold\Sigma_{\bold y|x}^{-1}(\bold y-\tilde{\bmath{\mu}}_{\bold y|x})\}\times\frac{g(x)}{g(y_1)}dxdy_2...dy_k.$$

The term in the exponent is:
$$-\frac{1}{2}\sum_{i=1}^k\left(\bold\Sigma_{\bold y|x}^{-1}\right)_{ii}\{(y_i-a'_i)-x\frac{\sigma_{xy_i}}{\sigma_{xx}}\}^2+$$
$$(-\frac{1}{2})\sum_{i\neq j}\left(\bold\Sigma_{\bold y|x}^{-1}\right)_{ij}\{(y_i-a'_i)-x\frac{\sigma_{xy_i}}{\sigma_{xx}}\}\{(y_j-a'_j)-x\frac{\sigma_{xy_j}}{\sigma_{xx}}\}$$
where $a'_i=a_i-\frac{E_g(X)}{\sigma_{xx}}\sigma_{xy_i}$.

We make the following substitutions:
$$(x,y_2,y_3,...,y_k)\longmapsto \bold y'=(y_1',y_2',y_3',...,y_k'),$$
$$y_1'=(y_1-a'_1)-x\frac{\sigma_{xy_1}}{\sigma_{xx}},$$
$$y_i'=(y_i-a'_i)-x\frac{\sigma_{xy_i}}{\sigma_{xx}},i=2,3,...,k.$$
Assuming that\ $\sigma_{xy_1}=Cov(X,Y_1)\neq 0$, the inverse map
$$\bold y'=(y_1',y_2',y_3',...,y_k')\longmapsto (x,y_2,y_3,...,y_k)$$
is given by:
$$x=\frac{\sigma_{xx}}{\sigma_{xy_1}}(y_1-y'_1-a'_1),$$
$$y_i=y'_i+a'_i+\frac{\sigma_{xy_i}}{\sigma_{xy_1}}(y_1-y'_1-a'_1),i=2,3,...,k,$$
$$\text{with Jacobian:}\ \ |\det\left(\frac{\partial(x,y_2,y_3,...,y_k)}{\partial(y_1',y_2',y_3',...,y_k')}\right)|=\frac{\sigma_{xx}}{\sigma_{xy_1}}.$$
The integrand becomes:
$$D \exp\{-\frac{1}{2}{\bold y'}^t\bold\Sigma_{\bold y|x}^{-1}\bold y'\}\left\{\frac{g\left(\frac{\sigma_{xx}}{\sigma_{xy_1}}(y_1-y'_1-a'_1)\right)}{g(y_1)}\right\}\frac{\sigma_{xx}}{\sigma_{xy_1}}.$$

By assumption,
$$\frac{g\left(\frac{\sigma_{xx}}{\sigma_{xy_1}}(y_1-y'_1-a'_1)\right)}{g(y_1)}\leq h(y_1)\,\,\text{for all}\,y_1,$$
for some non-negative function $h(\cdot)$  such that  $Eh(Z) <\infty$ when $Z$ has a Gaussian distribution.
We therefore have, by dominated convergence theorem
$$\lim_{y_1\to\infty}\int D \exp\{-\frac{1}{2}{\bold y'}^t\bold\Sigma_{\bold y|x}^{-1}\bold y'\}\left\{\frac{g\left(\frac{\sigma_{xx}}{\sigma_{xy_1}}(y_1-y'_1-a'_1)\right)}{g(y_1)}\right\}
\frac{\sigma_{xx}}{\sigma_{xy_1}}d\bold y'$$
$$=\int D \exp\{-\frac{1}{2}{\bold y'}^t\bold\Sigma_{\bold y|x}^{-1}\bold y'\}\lim_{y_1\to\infty}\left\{\frac{g\left(\frac{\sigma_{xx}}{\sigma_{xy_1}}(y_1-y'_1-a'_1)\right)}
{g(y_1)}\right\}\frac{\sigma_{xx}}{\sigma_{xy_1}}d\bold y'$$
$$=\int D \exp\{-\frac{1}{2}{\bold y'}^t\bold\Sigma_{\bold y|x}^{-1}\bold y'\}\lim_{y_1\to\infty}\left\{\frac{g\left(\frac{\sigma_{xx}}{\sigma_{xy_1}}(y_1-y'_1-a'_1)\right)}
{g(y_1-y'_1-a'_1)}\right\}\lim_{y_1\to\infty}\left\{\frac{g(y_1-y'_1-a'_1)}{g(y_1)}\right\}\frac{\sigma_{xx}}{\sigma_{xy_1}} d\bold y',$$
which, by our assumption on $g$, in turn equals
$$=\int D \exp\{-\frac{1}{2}{\bold y'}^t\bold\Sigma_{\bold y|x}^{-1}\bold y'\}\times\left(\frac{\sigma_{xy_1}}{\sigma_{xx}}\right)^\alpha\times\times 1\times \frac{\sigma_{xx}}{\sigma_{xy_1}}d\bold y'=\left(\frac{\sigma_{xy_1}}{\sigma_{xx}}\right)^{\alpha-1}\,\,.\,\,\Box$$

\bibliographystyle{abbrv}
\bibliography{project2011}

\begin{thebibliography}{10}

\bibitem{s:abe}
S.~Abe.
\newblock Axioms and uniqueness theorem for tsallis entropy.
\newblock {\em Physics Letters A}, 2000.

\bibitem{avel1}
M.~Avellaneda.
\newblock Minimum entropy calibration of asset-pricing models.
\newblock {\em International Journal of Theoretical and Applied Finance.},
  1:447--472, 1998.

\bibitem{avel3}
M.~Avellaneda, R.~Buff, C.~Friedman, N.~Grandchamp, L.~Kruk, and J.~Newman.
\newblock Weighted monte carlo: A new technique for calibrating asset pricing
  models.
\newblock {\em International Journal of Theoretical and Applied Finance.},
  4(1):91--119, March 2001.

\bibitem{black:litterman}
F.~Black and R.~Litterman.
\newblock Asset allocation: combining investor views with market equilibrium.
\newblock {\em Goldman Sachs Fixed Income Research}, 1990.

\bibitem{buch:kelly}
P.~Buchen and M.~Kelly.
\newblock The maximum entropy distribution of an asset inferred from option
  prices.
\newblock {\em The Journal of Financial and Quantitative Analysis.},
  31(1):143--159, March 1996.

\bibitem{cont:tankov}
R.~Cont and P.~Tankov.
\newblock Non-parametric calibration of jump-diffusion option pricing models.
\newblock {\em Journal of Computational Finance}, 7(3):1--49, 2004.

\bibitem{thomas:cover}
T.~Cover and J.~Thomas.
\newblock {\em Elements of Information Theory}.
\newblock John Wiley and Sons, Wiley series in Telecommunications, 1999.

\bibitem{csiszar1:1967}
I.~Csiszar.
\newblock Information-type measures of difference of probability distributions
  and indirect observations.
\newblock {\em Studia Scientifica Mathematica Hungerica.}, 2:299--318, 1967.

\bibitem{csiszar2:1967}
I.~Csiszar.
\newblock On topology properties of $f$-divergences.
\newblock {\em Studia Scientifica Mathematica Hungerica.}, 2:329--339, 1967.

\bibitem{csiszar:1972}
I.~Csiszar.
\newblock A class of measure of informativity of observation channels.
\newblock {\em Periodica Mathematica Hungerica.}, 2(1-4):191--213, 1972.

\bibitem{csiszar:1975}
I.~Csiszar.
\newblock I-divergence geometry of probability distribution and minimization
  problems.
\newblock {\em Annals of Probability}, 3(1):146--158, 1975.

\bibitem{csiszar:2008}
I.~Csiszar.
\newblock Axiomatic characterization of information measures.
\newblock {\em Entropy.}, 10:261--273, 2008.

\bibitem{dembo:zeit}
A.~Dembo and O.~Zeitouni.
\newblock {\em Large Deviations Techniques and Applications}.
\newblock Springer, Application of mathematics-38, 1998.

\bibitem{dos:santos}
R.~J.~V. dos Santos.
\newblock Generalization of shannon’s theorem for tsallis entropy.
\newblock {\em Journal of Mathematical Physics}, 21, 1997.

\bibitem{dup:eli}
P.~Dupuis and R.~Ellis.
\newblock {\em A Weak Convergence Approach to the Theory of Large Deviations}.
\newblock Wiley, Wiley series in probability and statistics, 1986.

\bibitem{feller:two}
W.~Feller.
\newblock {\em An Introduction to Probability Theory and its
  Applications,Vol-2}.
\newblock John Wiley and Sons Inc., New York, 1971.

\bibitem{cf1}
C.~Friedmann, J.~Huang, and S.~Sandow.
\newblock A utility-based approach to some information measures.
\newblock {\em Entropy.}, 9(1):1--26, 2007.

\bibitem{cf4}
C.~Friedmann, Y.~Zhang, and J.~Huang.
\newblock Estimating flexible, fat-tailed asset return distributions.
\newblock {\em
  \url{http://papers.ssrn.com/sol3/papers.cfm?abstract_id=1626342},}, 2010.

\bibitem{fritelli}
M.~Fritelli.
\newblock The minimal entropy martingale measures and the valuation problem in
  incomplete market.
\newblock {\em Mathematical Finance}, 10:39--52, 2000.

\bibitem{gibbs}
J.~Gibbs.
\newblock {\em Elementary Principles in Statistical Mechanics.}
\newblock New York: Scribner's, 1902, Reprint-Ox Bow Press, 1981.

\bibitem{glasserman:yu}
P.~Glasserman and B.~Yu.
\newblock Large sample properties of weighted monte carlo estimators.
\newblock {\em Operation Research}, 53(2):298--312, 2005.

\bibitem{goll}
T.~Goll and L.~Ruschendorf.
\newblock Minimal distance martingale measure and optimal portfolios consistent
  with observed market prices.
\newblock In {\em Stochastic Monographs}, volume~12, pages 141--154. Taylor and
  Francis, London, 2002.

\bibitem{golshani:pasha}
L.~Golshani, E.~Pasha, and Y.~Gholamhossein.
\newblock Some properties of renyi entropy and renyi entropy rate.
\newblock {\em Information Sciences}, 170:2426--2433, 2009.

\bibitem{jns:one}
E.~Jaynes.
\newblock Information theory and statistical mechanics.
\newblock {\em Physics Reviews.}, 106:620--630, 1957.

\bibitem{jns:two}
E.~Jaynes.
\newblock {\em Probability Theory: The Logic of Science}.
\newblock Cambridge University Press, 2003.

\bibitem{fq}
M.~Jeanblanc, S.~Kloeppel, and Y.~Miyahara.
\newblock Minimal $f$-$q$ martingale measures for exponential levy processes.
\newblock {\em Annals of Applied Probability}, 17(5/6):1615--1638, 2007.

\bibitem{jizba:arimitsu}
P.~Jizba and T.~Arimitsu.
\newblock The world according to renyi:thermodynamics of multi fractal systems.
\newblock {\em Annals of Physics}, 312:17--59, 2004.

\bibitem{kallsen}
M.~Kallsen.
\newblock Utility-based derivative pricing in incomplete markets.
\newblock In {\em Mathematical Finance - Bachelier Congress 2000}, pages
  313--338. Springer, Berlin, 2002.

\bibitem{jn:kapur}
J.~Kapur.
\newblock {\em Maximum Entropy Models in Science and Engineering}.
\newblock New Age International Publishers, Wiley series in Telecommunications,
  2009.

\bibitem{khin:chin}
A.~I. Khinchin.
\newblock {\em Mathematical foundation of information theory}.
\newblock Dover, New York, 1957.

\bibitem{kitamura:stutzer}
Y.~Kitamura and M.~Stutzer.
\newblock Entropy-based estimation methods.
\newblock In {\em Encyclopedia of Quantitative Finance}, pages 567--571. Wiley,
  2010.

\bibitem{Luenberger2}
D.~Luenberger.
\newblock {\em Linear and Nonlinear Programming 2nd Edition}.
\newblock Springer, 2003.

\bibitem{fully:flexible}
A.~Meucci.
\newblock Fully flexible views: theory and practice.
\newblock {\em Risk}, 21(10):97--102, 2008.

\bibitem{mina:xiao}
J.~Mina and J.~Xiao.
\newblock Return to riskmetrics: the evolution of a standard.
\newblock {\em RiskMatrics publications}, 2001.

\bibitem{least:disc}
J.~Pazier.
\newblock Global portfolio optimization revisited: a least discrimination
  alternative to black-litterman.
\newblock {\em ICMA Centre Discussion Papers in Finance}, July 2007.

\bibitem{nr}
W.~Press, S.~Teukolsky, W.~Vetterling, and B.~Flannery.
\newblock {\em Numerical Recipes 3rd Edition: The Art of Scientific Computing}.
\newblock Cambridge University Press, 2007.

\bibitem{qian:gorman}
E.~Qian and S.~Gorman.
\newblock Conditional distribution in portfolio theory.
\newblock {\em Financial analyst journal.}, 57(2):44--51, March-April 1993.

\bibitem{alfred:renyi}
A.~Renyi.
\newblock On measures of entropy and information.
\newblock In {\em Proceedings of the 4th Berkeley Symposium on Mathematics,
  Statistics and Probability}, pages 547--561, 1960.

\bibitem{Shannon:one}
C.~Shannon.
\newblock A mathematical theory of communication.
\newblock {\em The Bell System Technical Journal}, 27:379--423,623--656, 1948.

\bibitem{zasta}
W.~Slomczynski and T.~Zastawniak.
\newblock Utility maximizing entropy and the second law of thermodynamics.
\newblock {\em The Annals of Probability.}, 32(3A):2261--2285, 2004.

\bibitem{stutzer}
M.~Stutzer.
\newblock A simple non-parametric approach to derivative security valuation.
\newblock {\em Journal of Finance}, 101(5):1633--1652, 1997.

\bibitem{tsallis:one}
C.~Tsallis.
\newblock Possible generalization of boltzman gibbs statistics.
\newblock {\em Journal of Statistical Physics}, 52:479, 1988.

\bibitem{cf3}
X.~Zhou, J.~Huang, C.~Friedmann, and S.~Sandow.
\newblock Private firm default probabilities via statistical learning theory
  and utility maximization.
\newblock {\em Journal of Credit Risk.}, 2(1):1--26, 2006.

\end{thebibliography}

\end{document}